\documentclass[aps,prb,twocolumn,superscriptaddress]{revtex4-1}
\usepackage{graphicx}
\usepackage{amsmath}

\begin{document}
\title{A combined first-principles and thermodynamic approach to M-Nitronyl Nitroxide (M=Co, Mn) spin helices} 


\author{Marco Scarrozza}
\affiliation{CNR SPIN, UOS L'Aquila, via Vetoio 10, 67100 Coppito, L'Aquila, Italy}
\email[]{marco.scarrozza@spin.cnr.it}
\author{Alessandro Vindigni}
\affiliation{Laboratorium f\"ur Festk\"orperphysik, ETH Z\"urich, CH-8093 Z\"urich, Switzerland}
\email[]{vindigni@phys.ethz.ch}
\author{Paolo Barone}
\affiliation{CNR SPIN, UOS L'Aquila, via Vetoio 10, 67100 Coppito, L'Aquila, Italy}
\author{Roberta Sessoli}
\affiliation{Laboratory of Molecular Magnetism (LaMM), Dip.to di Chimica Ugo Schiff, Universit\`a degli Studi di Firenze \& INSTM RU, via della Lastruccia 3-13, 50019 Sesto Fiorentino (FI), Italy.}
\author{Silvia Picozzi}
\affiliation{CNR SPIN, UOS L'Aquila, via Vetoio 10, 67100 Coppito, L'Aquila, Italy}
\date{\today}

\begin{abstract}
The properties of two molecular-based magnetic helices, composed of 3$d$ metal Co and Mn ions bridged by Nitronyl Nitroxide radicals, are investigated by density functional calculations. 
Their peculiar and distinctive magnetic behavior is here elucidated by a thorough description of their magnetic, electronic, and anisotropy properties.
Metal ions are antiferromagnetically coupled with the radicals, leading to a ferrimagnetically ordered ground state. 
A strong metal-radical exchange coupling is found, about 44 meV and 48 meV for Co- and Mn-helices, respectively. 
The latter have also relevant next-nearest-neighbor Mn-Mn antiferromagnetic interactions (of $\sim$ 6 meV).
Co-sites are characterized by non-collinear uniaxial anisotropies, whereas Mn-sites are rather isotropic.
A key result pertains to the Co-helix: the microscopic picture resulting from density-functional calculations allows us to propose a spin Hamiltonian of increased complexity 
with respect to the commonly employed Ising Hamiltonian, suitable for the study of finite-temperature behavior,
and that seems to clarify the puzzling scenario of multiple characteristic energy scales observed in experiments. 
\end{abstract}


\maketitle


\section{Introduction\label{Introduction}}
Magnetic bistability at the molecular level has attracted broad
interest over the last decades in view of magnetic-storage applications\cite{gatteschi2006,Bartolome2014}. 
From a fundamental perspective, molecules
carrying a small number of interacing paramagnetic centers but behaving
like bulk magnets, otherwise known as Single-Molecule Magnets (SMMs),
allowed the direct observation of quantum tunneling of the magnetization\cite{sessoli1993},
of the Berry phase\cite{Wernsdorfer_Science99} and magnetic chiral
degrees of freedom\cite{LuzonPRL2008} (just to cite few remarkable
phenomena) as well as the manipulation of magnetization by light irradiation\cite{Sorace_PRB_03}
or via an applied voltage\cite{Bogani_NatMat08}. A large easy-axis
magnetic anisotropy is a key-requirement to achieve bistability (namely,
blocking of the magnetization) in molecular magnets and is provided
by metal ions with unpaired electrons  carrying a finite total angular momentum.
These anisotropic building
blocks are coordinated to organic ligands that are functionalized
in order to favor or shield the propagation of exchange interactions.
This synthetic strategy naturally leads to the formation of magnetic
systems with reduced dimensionality\cite{bogani2008,gatteschi2013},
in which exchange paths may have finite connectivity (like in SMMs)
or create one- (1D) or two-dimensional networks. 

During the 70s and the 80s, molecular 1D systems have been the object of intense investigation
and helped testing the basic principles of \emph{equilibrium} statistical
physics on realistic systems\cite{SVW_AdvPhys76,DeJongh-Miedema_AdvPhys01}.
The natural evolution of this research line led to study quantum-phase
transitions, which are still actively explored on 1D magnetic systems\cite{Sachdev_Science_00,Coldea_Science_10,Simon_Nature_11}.
Starting from 2001, the observation of slow relaxation of the magnetization
in molecular spin chains promoted them to the role of prototypical
systems for the study of \emph{out-of-equilibrium} phenomena as well.
By analogy with SMMs, slow-relaxing spin chains were called Single-Chain Magnets (SCMs)\cite{Clerac2002}. 
However, the basic properties of SCMs are only partially
similar to SMMs, since in the first ones slow relaxation is not exclusively
determined by the total magnetic anisotropy. In fact, in SCMs this
phenomenon may be associated either with the development of short-ranged
spin-spin correlations upon cooling or with the nucleation of a domain
wall (DW) at a defect site\cite{Coulon06Springer,BoganiPRL2004,CoulonPRB04,VindigniAPL05}.

 SCM behavior has been first observed in the compound Co(hfac)$_{2}$(NITPhOMe),
hereafter called CoPhOMe, which is composed of Co(hfac)$_{2}$ moieties
bridged by NITPhOMe radicals (where hfac=hexafluoroacetylacetonate
and NITPhOMe=4'-methoxy-phenyl-4,4,5,5-tetramethylimidazoline-1-oxyl-3-oxide)
arranged in chiral 1D arrays\cite{caneschi2001}. Then, slow relaxation
of the magnetization and hysteresis effects observed in CoPhOMe were
rationalized in terms of the kinetic Ising model proposed by Glauber\cite{glauber1963}.
In the comparatively large literature on SCMs that followed, slow
relaxation has essentially always been interpreted in terms of the
Glauber model (or variations of it), which
indeed allows to explain several remarkable features of CoPhOMe\cite{caneschi2001,Coulon_PRL09,BoganiPRL2004,CoulonPRB07,CoulonPRB04,VindigniAPL05,VindigniJPCM09,PiniPRB11,Coulon06Springer,Bogani_NatMat13}. 
Within the Ising model  both the correlation length and the relaxation time 
are predicted to diverge at low temperature according to an Arrhenius law and with the \textit{same}  
energy barrier. However, experimentally these two energy scales turn out to be different in CoPhOMe\cite{caneschi2001}, thus calling for 
a deeper microscopic understanding of electronic and thermodynamic properties of this compound.  
In the present study, this issue is addressed showing that the mismatching between the two energy scales mentioned above 
can be justified relaxing the hypothesis of  large (virtually
infinite) uniaxial magnetic anisotropy, underlying the description in terms of the Ising model.  

According to the present understanding, the emergence of SCM behavior
in a given molecular compound requires that \emph{i}) some anisotropy
prevents the magnetization from reorienting easily, \emph{ii}) the
exchange coupling mostly develops along one dimension and \emph{iii})
the relaxation time of the magnetization becomes macroscopic well
above the temperature at which residual 3D interactions trigger magnetic
ordering. Regarding the first requirement, high-spin $d^7$ cobalt(II)
ions in distorted octahedral environment are the source of anisotropy
in CoPhOMe. The 1D character (\emph{ii}) is guaranteed by the strong
metal-radical antiferromagnetic exchange along the chain and by the
absence of an exchange path between magnetic centers belonging to
different chains\cite{caneschi2001}. The (residual) dipolar interaction
indeed couples also spins belonging to different chains of the crystal,
but it is considered to be too small and partially frustrated to induce 3D
magnetic order in CoPhOMe at relatively high temperatures. Interestingly,
its isostructural Mn-based compound,  Mn(hfac)$_{2}$(NITPhOMe) (MnPhOMe
in the following),
comprising isotropic high-spin $d^5$ ions,
does not show slow relaxation of the magnetization
but  exhibits a transition to a magnetically ordered phase at low
temperature (4.8 K) \cite{caneschi1991} . These isostructural molecular
helices have allowed a comparative investigation of the interplay
between structural chirality and magnetism, evidencing a giant magneto-chiral
dichroism in the hard X-ray range for the Co-based noncollinear
spin chain, absent in the isotropic manganese analogue \cite{Sessoli_NatPhys2014}.
Moreover visible light has been found to promote fast relaxation
in CoPhOMe through a kick-off mechanism for the nucleation of domain walls\cite{Bogani_NatMat13}.

A comparative first-principles study of these archetypical spin chains
is therefore of great relevance and may  evidence which basic ingredients
are needed to produce slow dynamics avoiding -- at the same time --
3D magnetic ordering. As for the last requirement (\emph{iii}), from
the microscopic point of view a crucial issue is the relative strength
of magnetic anisotropy and characteristic exchange energy, the former
being also a key requirement for the observation of magneto-chiral
effects. Here, we address this issue by studying the electronic and
magnetic properties of both CoPhOMe and MnPhOMe via density functional
calculations. From these, an estimate for the intrachain exchange
coupling among different magnetic centers is provided as well as for
the magnetic anisotropy on the magnetic ions. The $ab-initio$
analysis, then, allows us to propose an effective, albeit realistic,
spin model suitable to study the thermodynamic properties as well
as slow relaxation effects of CoPhOMe. Remarkably, our results suggest
that the exchange interactions in this SCM are larger than what previously
believed and represent the largest energy scale in the model, implying
the emergence of broad DWs and thus directly affecting the slow relaxation
properties of the SCMs; moreover, the increased complexity of our
proposed thermodynamic model can qualitatively explain the observation
of multiple energy scales in CoPhOMe. 

The paper is structured as follows. 
In the first part (Sec.~\ref{DFT}), the description of the microscopic properties of the systems investigated by density functional calculations is presented.  
After a brief methodological section reporting the technical details of the calculations (Sec.~\ref{Methods}), 
the structural properties of the magnetic chains are reviewed, with particular attention given to the magnetic building units (Sec.~\ref{Structural properties}).
A detailed analysis of the electronic properties is given in Sec.~\ref{Electronic properties},
followed by a discussion of the magnetic properties of two chains (Sec.~\ref{Magnetic properties}).
The study of the energetics of several higher-energy spin configurations allowed to identify and estimate the exchange interactions at play,
whereas the magnetic anisotropy has been quantified in related mononuclear Co- and Mn-complexes.  
In the second part (Sec.~\ref{Thermodynamics}), the results obtained by density functional calculations 
are used as input for an atomistic spin Hamiltonian through which the basic thermodynamics of CoPhOMe can be described.  
Finally, the main results are summarized in Sec.~\ref{Summary}.

\section{\label{DFT}Microscopic properties by first-principles}   
\subsection{\label{Methods}Methods}
We performed \emph{ab-initio} calculations within spin-density functional theory in the generalized gradient approximation\cite{perdew1996}
as implemented in the VASP materials modelling package\cite{kresse1996, kresse1999}; 
for Co and Mn, the $3d4s$ states were treated as valence states, while for C, N, O, and F, the $2s2p$ were considered as valence states,
within the projector-augmented waves (PAW) method\cite{blochl1994}. 
Magnetic calculations were performed both in the  collinear and non-collinear formalism\cite{hobbs2000}, including spin-orbit interaction when required.
A plane-wave cut-off of 400 eV and a $2\times2\times1$ k-points grid in the primitive cell's Brillouin zone were adopted, 
and energy convergence was ensured by a strict threshold of 10$^{-6}$ eV. 
To model molecular systems comprising transition metals properly, an accurate treatment of electronic correlation is crucial. 
We have addressed this aspect by performing extensive calculations within the GGA+U approach in the Liechtenstein formalism\cite{liechtenstein1995},
in which a set of two effective parameters (U,J) is employed treat electron-correlation, where U describes the on-site Coulomb repulsion and J is the on-site exchange-correction;
calculations were also performed in the simplified Dudarev's scheme\cite{dudarev1998}. 
For each system investigated, a systematic study of the magnetic properties was carried out, spanning U and J parameters over a large range of values. In this way, we provided an overview of the dependence of the magnetic properties on the correlation parameters,  
and -- most importantly -- we ensured that our results were robust and conclusions were not biased by an inappropriate model description.
Unless otherwise specified, results for three sets of (U,J) values will be presented in the following, denoted as:
set \emph{a} (U$_{M}$ = 5 eV); set \emph{b} ( U$_{M}$ = 6.5 eV; J$_{M}$ = 1.5 eV)  \emph{i.e.} an \emph{effective} U as in set \emph{a}
but with the inclusion of a sizeable J value, which turned to be  crucial to the aim of modeling the magnetic anisotropy;
and finally, a set \emph{c} (U$_{M}$ = 6.5 eV; J$_{M}$ = 1.5 eV; U$_{O}$=U$_{N}$ = 6 eV)
that enhances the degree of localization by inclusion of a Hubbard-U also on $p$-states of O and N atoms.

\begin{figure}[ht]
\begin{center}
\includegraphics[width=8.8cm]{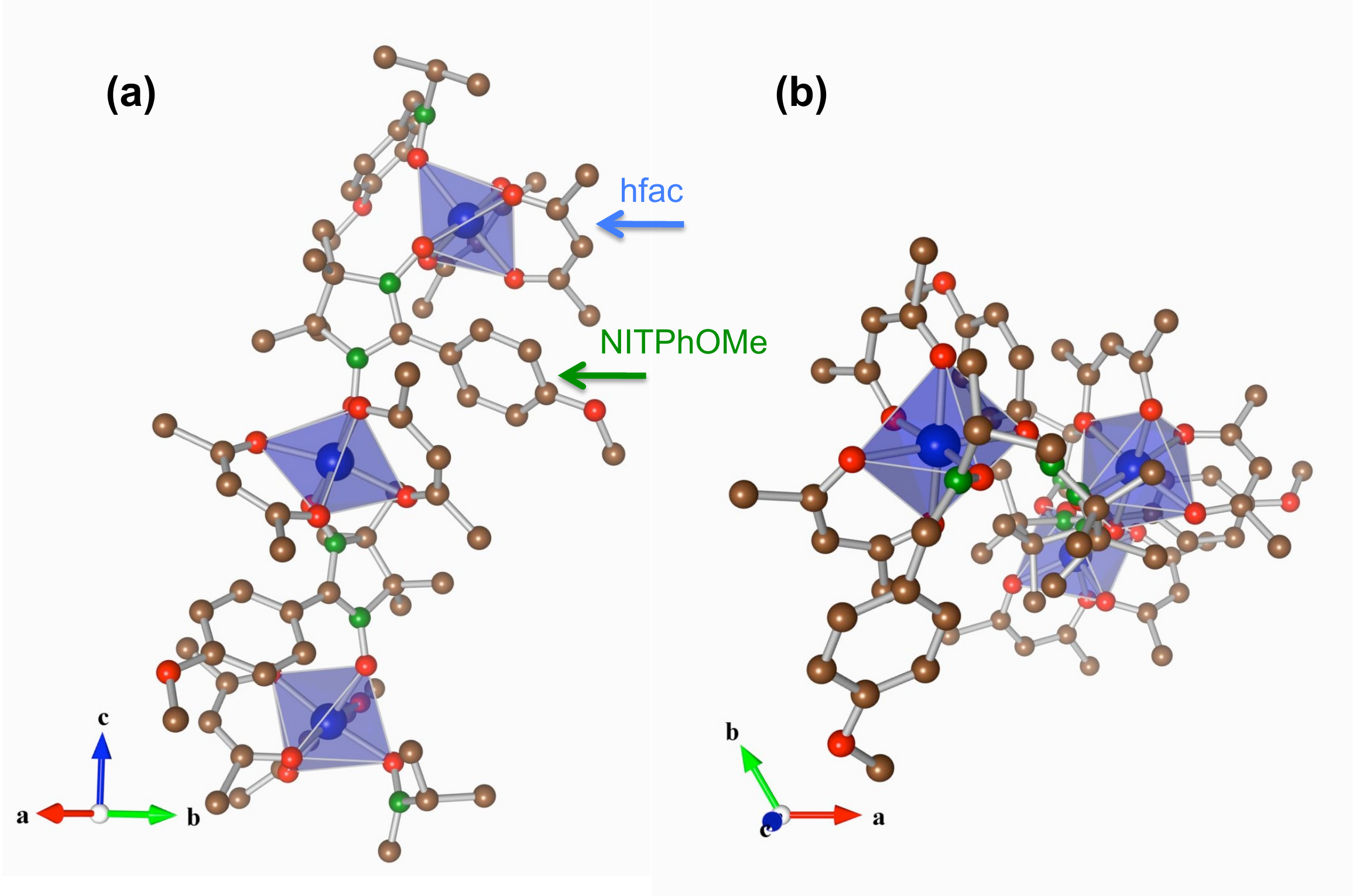}
\caption{\label{Fig1}Color online. Crystal structures of isostructural CoPhOMe and  MnPhOMe, (a) side and (b) top views.
The M (Co, Mn) atoms are depicted as large blue spheres, and the O, N, C atoms are red, green and brown spheres, respectively. For the sake of clarity, F and H atoms are not shown.
The hfac and NITPhOMe molecules are highlighted in the side view; the trigonal symmetry of the chains can be recognized from the top view.}
\end{center}
\end{figure}

\begin{figure}
\includegraphics[width=1\linewidth]{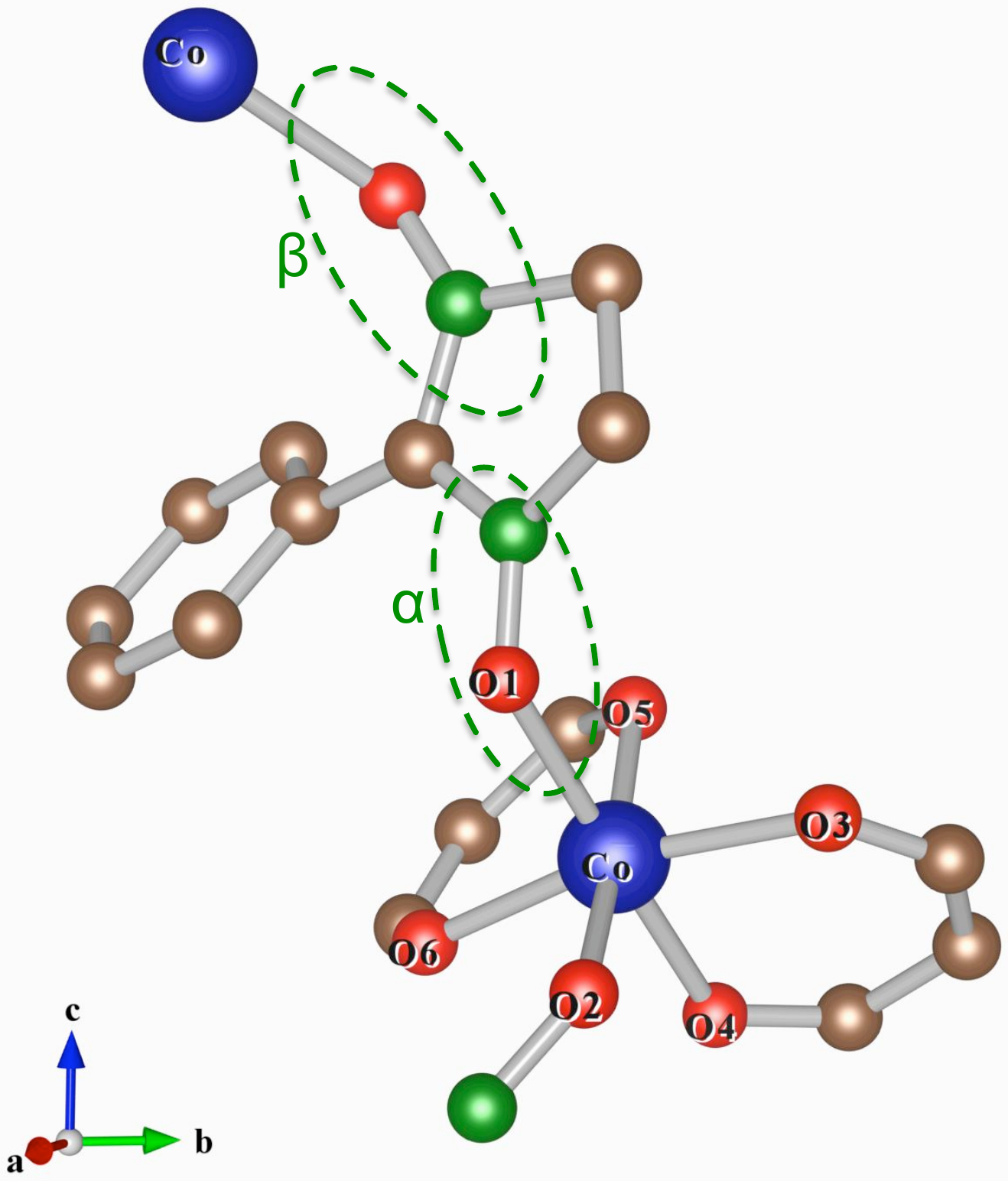}
\caption{\label{Fig2} Color online. 
Magnified view of the molecular cluster of CoPhOMe, comprising the Co-octahedral complex and the NITPhOMe radical bridging two Co-atoms. 
A similar structure characterizes also MnPhOMe.} 
\end{figure}

\subsection{Structural properties\label{Structural properties}}
The crystal structures of the two chains in shown Fig.~\ref{Fig1}, 
composed of alternating M(hfac)$_{2}$ moieties and NITPhOMe (denoted also with the shortened term NIT, in the following) 
organic radicals arranged in 1D arrays with helical structures generated by a three-fold screw axis.  Despite the absence of chiral constituents the compounds form enantiopure crystals, crystallising either in the chiral  $P3_{1}$ or  $P3_{2}$ space groups. 
 The experimental structures were used for the calculations,
with lattice parameters: $a=b=11.294$ {\AA}, $c=20.570$ {\AA} for the Co-chain; and $a=b=11.281$ {\AA}, $c=20.846$ {\AA} for the Mn-chain. 

We now discuss in closer detail the structural features of the magnetic bricks of the chains, starting with CoPhOMe. A view of the molecular cluster is shown in Fig.~\ref{Fig2}.
The Co-ion is in a distorted octahedral environment, coordinated by the O atoms to two hfac ligands (O3 and O4, and O5 and O6 for the two molecules, see Fig.~\ref{Fig2}), while the remaining sites host the terminating O atoms of the NITPhOMe radicals in \emph{cis} position, that connect two consecutive Co(hfac)$_2$ moieties in a chain.
The Co-octahedral complex has low symmetry, lacking a clear regular pattern. Nevertheless, as an attempt to rationalize it,
a main tetragonal compression of the octahedron can be distinguished along the axis connecting the O atoms of the two hfac molecules, 
the O3 and O6 sites denoted as \emph{apical}, whose distance is the shortest one (4.046 {\AA}).
This compression indeed corresponds to the maximum distortion from the octahedral symmetry observed in the O3-Co-O6 angle, which is 166.62$^{\circ}$, 
while the angle formed by the two edging O atoms of the cis-coordinated NIT-radicals O1-Co-O2 (cfr.  Fig.~\ref{Fig2}) is 85.25$^{\circ}$.   
Next, a secondary rhombic-like distortion on the equatorial plane of the polyhedron can be revealed by the different distances between the vertical O atoms, 
namely 4.147 {\AA} for O2-O5 and 4.190 {\AA} for O1-O4.
Focusing on the bridging NITPhOMe radicals, in fact two moieties can be distinguished as shown in Fig.~\ref{Fig2}, denoted as $\alpha$ and $\beta$ moieties.
These are characterized by non-equivalent structural patterns, whereby Co-O bonds are slightly larger in $\alpha$ (Co-O1=2.108 {\AA}) than $\beta$ (Co-O2=2.097 {\AA})
by $\sim$ 0.5\%,
and viceversa the O-N and N-C bonds are shorter in $\alpha$ respect to $\beta$ moiety, by 1\% and 2\% respectively. 

Moving to MnPhOMe, the Mn-octahedral complex shows a similar distortion pattern to the Co-counterpart, characterized by a main tetragonal axial compression, 
with apical O atoms  distances of 4.026 \AA\ , and a smaller distortion on the equatorial plane (trans O's distances are 4.326 and 4.252 \AA\ ). 
The NITPhOMe features a similar structural pattern to the one discussed above for the Co-case, 
whereby the Mn-O bonds are larger in the $\alpha$ (2.143 \AA) with respect to the $\beta$ moiety (2.121 \AA) by 1\%,
whereas the O-N and N-C bonds are shorter in the $\alpha$ moiety by 2.5\% and 0.1\% compared to $\beta$.
Globally, Co-O bond lengths vary in the range (2.03  $\textendash$ 2.11 \AA), whereas Mn-O ones are in the range (2.09 $\textendash$ 2.19 \AA),
with Mn-O bonds longer than Co-O ones by about 3 $\textendash$ 4\%.

\begin{figure}
\includegraphics[width=1\linewidth]{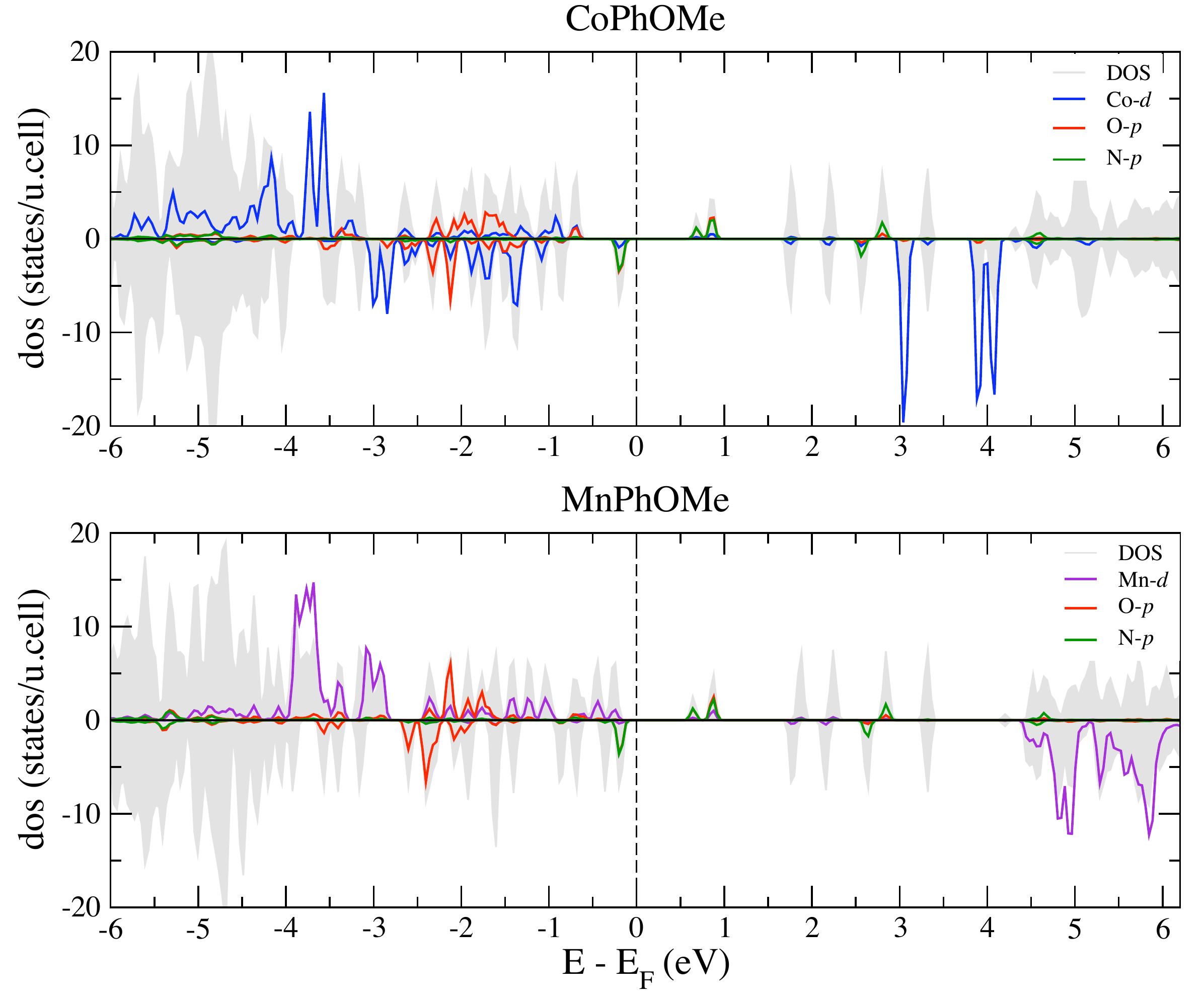}  
\caption{\label{DOS}Color online. Computed density of electronic states (DOS) of CoPhOMe (upper panel) and MnPhOMe (lower panel).
Total DOS are shown (rescaled by a factor 0.1 for the sake of clearness) together with the projected DOS 
relative to metal (Co, Mn) $d$-states, and $p$-states of selected O and N atoms of the NITs.}
\end{figure}

\subsection{\label{Electronic properties}Electronic properties}
Experimentally, CoPhOMe is described by Co(II) ions in the high-spin 3/2 state, and NITPhOMe radicals carrying a free electron (spin 1/2) that mediates the exchange interaction between the Co ions. 
As already stated, Co ions are characterized by a large easy-axis magnetic anisotropy \cite{caneschi2001,gatteschi2013}.  
Aiming at a systematic analysis of the dependence of magnetic properties on the correlation parameters, we initially included the Hubbard-U parameter in the $d$-Co states and varied it in the energy interval (0 $\textendash$ 5) eV, by searching for the magnetic ground state.
We found that the inclusion of U$_{Co}$ values at least of 3 eV is needed to correctly reproduce the high spin state of Co(II). 
As expected, the local magnetic moment on Co enhances at higher U values; an optimal value of U$_{Co}$=5 eV was chosen.
Let us briefly discuss the computed electronic properties of the magnetic chains, and their implications for the magnetic properties.
In Fig.~\ref{DOS} the total density of electronic states (DOS), together with the projected DOS of the magnetic elements of the chains,
the M-ions and the O and N-sites of the radicals, are shown. These calculations refer to the (U,J)-set $a$.
For these U values, the chains are small gap insulators, with an HOMO-LUMO gap that amounts to 0.60 eV in both Mn and Co cases.   
We however caution the reader that the quantitative estimate of the gap value could be strongly affected by the exchange-correlation functional,
as the starting GGA-functional is known to severely underestimate the energy position of the excited states. 
The HOMO and LUMO states are mostly composed of the $p$-states of O and N of the NITs with equal weight,
while the M-states are pretty far from the gap region.
In CoPhOMe, the high-spin (S=3/2) electronic configuration is here visible in the $d$-resolved Co-states:
the spin-up channel is fully occupied, the unoccupied states in the spin-down channel are resolved as three peaked states,
a $t_{2g}$ state at 3 eV above the Fermi level (E$_{F}$), and two $e_{g}$ states at higher energy ($\sim$ 4 eV above E$_{F}$).
The occupied $d$-states strongly hybridize with the O-$p$ states of the NITs in the energy range between 3 and 1 eV below the E$_{F}$.
The DOS of MnPhOMe presents similar features to the Co-case, with the characteristic composition of HOMO and LUMO states by O and N atoms of the bridging radicals. 
The main difference is found for the unoccupied down-spin states which appear at higher energy, as they start at 4.5 eV above the E$_{F}$,
where two major peaks can be distinguished, composing the $t_{2g}$ states at 5 eV, followed by the $e_{g}$ states at 6 eV.      
Also in this case the occupied Mn-$d$ states are hybridized with the O-$p$ states.

\subsection{\label{Magnetic properties}Magnetic properties}
\subsubsection{Magnetic moments}
The converged magnetic ground state of CoPhOMe has a ferrimagnetic (FiM) order, in which the neighboring Co-atoms and NITPhOMe molecules are coupled antiferromagnetically (AF), 
and well reproduces the picture of a Co-ion in high spin state and a free electron delocalized on the bridging NITs.
A total magnetic moment of 6 ${\mu_{B}}$/unit cell is found for this FiM state,
as expected since the cell can be seen as containing  three AF coupled (Co-NIT) magnetic dimers,
each one featuring an effective moment of 2 ${\mu_{B}}$.  
In Tab.~\ref{MagMoments}, the values of the projected magnetic moments of the Co-ions as well as of the NITs are reported. 
The Co site has a local moment of 2.72 ${\mu_{B}}$, and the free electron in the bridging NITs is localized on the O-N groups, 
with a slight unbalance of local magnetic moment towards the O-N group in the ${\beta}$ configuration (0.04 ${\mu_{B}}$) with respect to the ${\alpha}$ one;
the C atom is weakly AF coupled to the two O-N groups.

Turning to the other chain, in which the Mn(II) ions are in $3d^{5}$ electronic configuration, a FiM ground state was also found, given by the AF coupling of the magnetic bricks Mn and NITs, and a total magnetic moment amounting to 12 ${\mu_{B}}$/unit cell. More precisely, the projected magnetic moments are 4.59 ${\mu_{B}}$ for the Mn-ions;
the unbalance between local magnetic moments ${\alpha}$ versus ${\beta}$ on the O-N groups,
as discussed above for the Co-counterpart, here is reduced, with moments of $-0.16$ and $-0.18$ ${\mu_{B}}$ on the O and N atoms, respectively.  

\subsubsection{Spin density}
\begin{figure}
\includegraphics[width=1\linewidth]{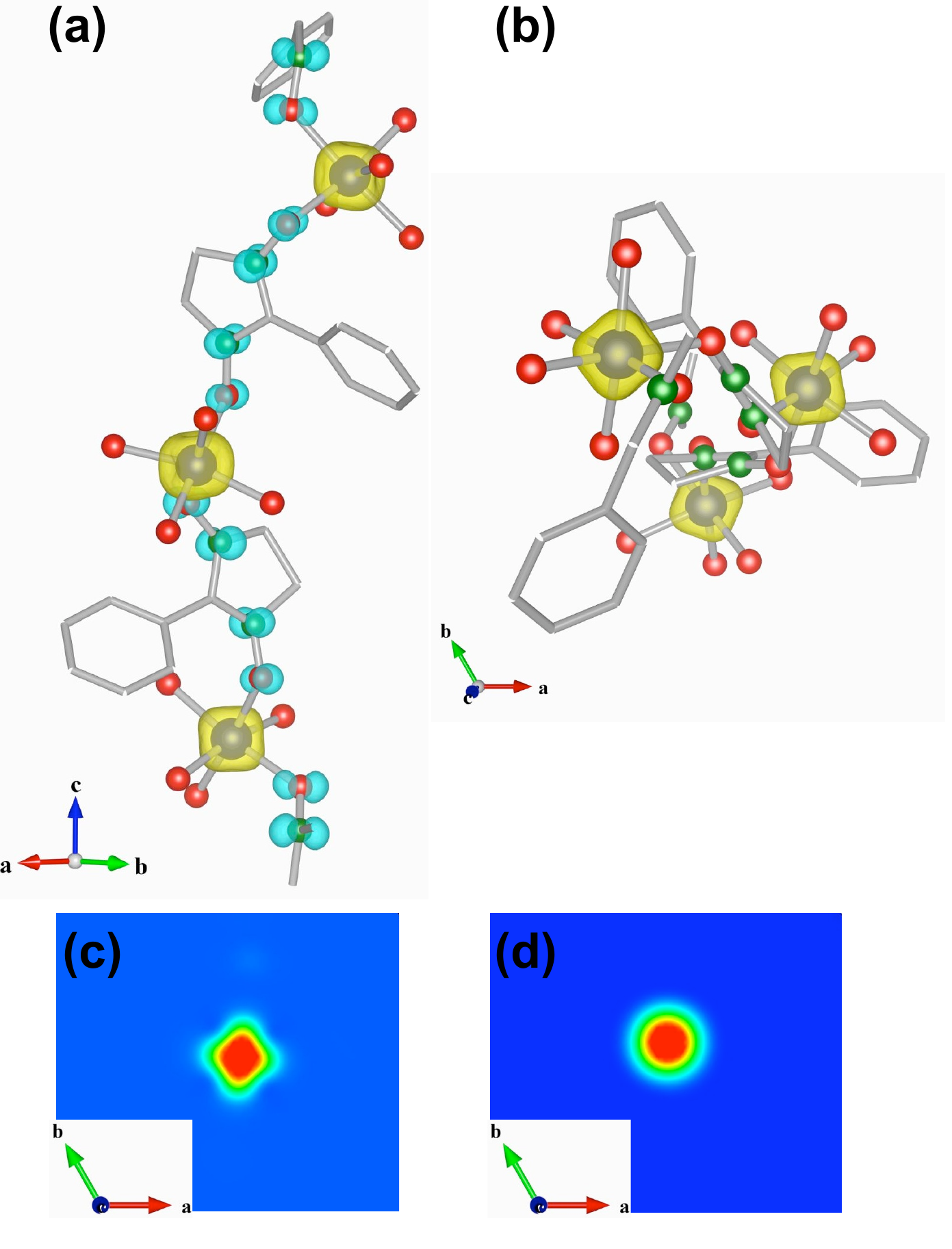}
\caption{\label{SpinDensity}Color online. 
Isosurface plots of the spin density of CoPhOMe: a side view in (a), and top view in (b), the latter displaying only the positive spin density for the sake of clearness.
Positive (negative) values of the spin density are depicted as yellow (cyan) lobes.
2D section cuts of the spin density taken in a plane parallel to $ab$ and centered on M-ions, for CoPhOMe in (c) and MnPhOMe in (d).}      
\end{figure}

A complementary detailed picture of the magnetic structure of the CoPhOMe is provided by the spin density isosurfaces  shown in Fig.~\ref{SpinDensity},
with positive spin density centered on Co-sites, whereas the negative spin density (the radical's free electron) is equally distributed on 
the $p$-states of the O and N atoms of the NITs (cfr. Fig.~\ref{SpinDensity}(a)).
The MnPhOMe is characterized by a similar spin density distribution (not shown).
The only difference that can be discerned between the two chains pertains the shape of the spin density localized on the M-ions, that mirrors their electronic structure.
As a matter of fact, the Mn-ion are characterized by a perfectly spherical distribution of the spin density, as shown in the 2D sections cuts (Fig.~\ref{SpinDensity}(d)) 
due to the full (zero) occupancy of the spin-up (down) channel.
On the other hand, the spin density on the Co-sites shows distinctive features of partially occupied $e_{g}$ orbitals,
whereby the lobes directed towards the O's of the octahedral complex can be distinguished in Fig.~\ref{SpinDensity}(c).   
Moreover, the non-collinearity of local anisotropies of the Co-octahedral complexes can be seen 
in Fig.~\ref{SpinDensity}(b), each showing features which are rotated by 120$^{\circ}$ with respect to the preceding Co-sites by trigonal symmetry.    
  
\begin{table}
\caption{\label{MagMoments}Projected magnetic moments on M (Co, Mn) and on the O-N-C-N-O group of the NITPhOMe radicals 
corresponding to the FiM magnetic ground state, for CoPhOMe and MnPhOMe. 
The atomic projections are performed within spheres of radius as follows: Co (1.30 {\AA}), Mn (1.32 {\AA}), O (0.82 {\AA}), N (0.74 {\AA}), and C (0.86 {\AA}).  
Values are in ${\mu_{B}}$.} 
\begin{ruledtabular}
\begin{tabular}{lccccc}
M 		  &	O$_{\alpha}$    &	N$_{\alpha}$	&	C		& N$_{\beta}$	&	O$_{\beta}$  \\ \hline
Co (2.72) &	$-0.16$		&	$-0.18$		&	0.05		& $-0.20$	&		$-0.17$	\\			
Mn (4.59) &	$-0.16$		&	$-0.18$		&	0.05		& $-0.18$	&		$-0.16$	\\
\end{tabular}
\end{ruledtabular}
\end{table}
 
\begin{figure}
\includegraphics[width=1\linewidth]{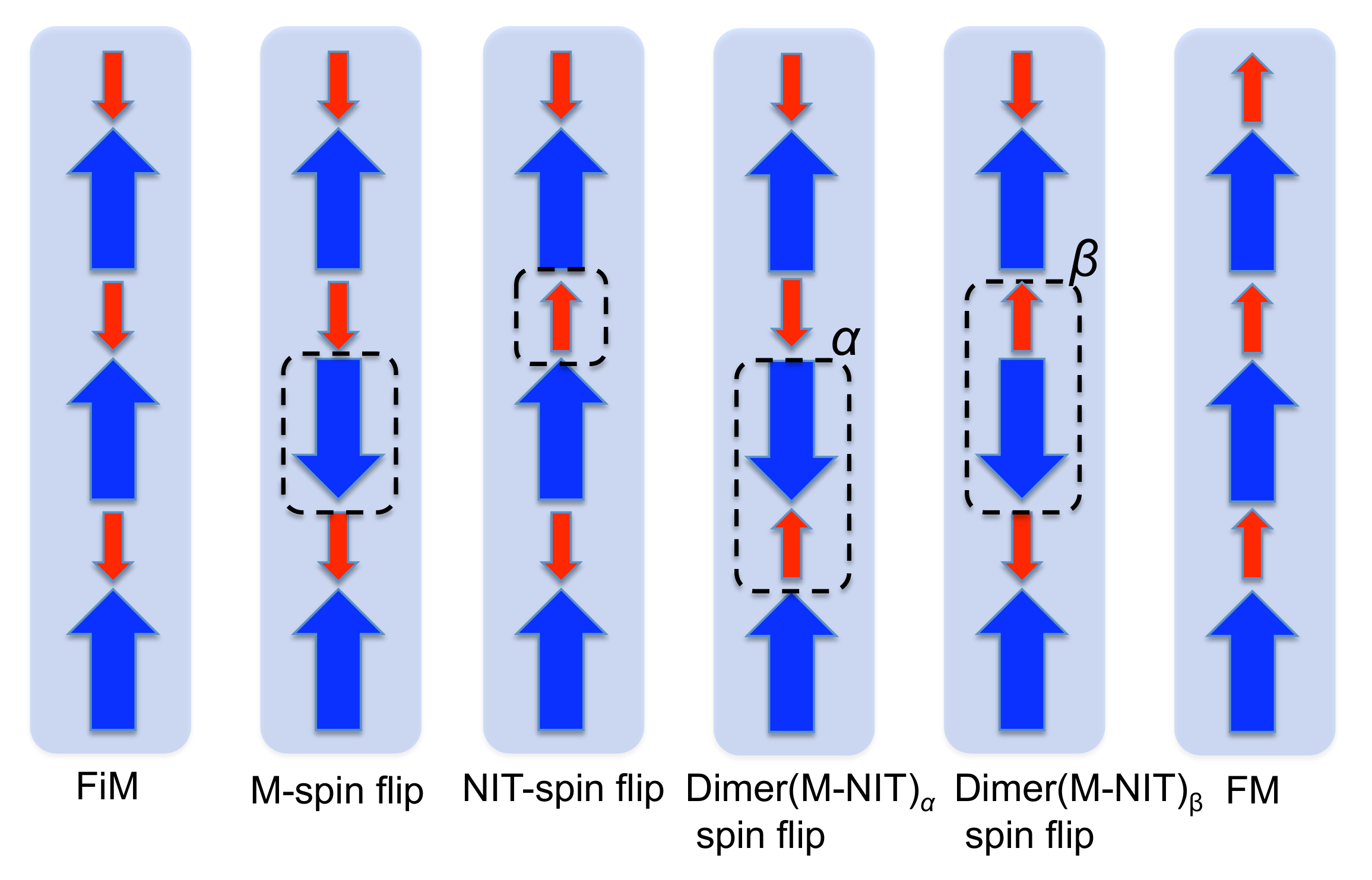}
\caption{\label{MagConfigs}Color online. Schematic view of the set of collinear magnetic configurations modeled 
to calculate the exchange interactions in MPhOMe chains, with M = Co, Mn.
From left to right: the FiM (ground state), the single M- and NIT-spin flips, the dimer (M-NIT) spin-flips at the $\alpha$ and $\beta$ bondings,
and finally the FM magnetic state. The M- and NIT-spins are depicted as large blue and small red arrows, respectively.}
\end{figure}

\subsubsection{Exchange coupling constants}
In order to estimate the magnetic interactions at play in the chains, 
spin-polarized density functional calculations were performed for an extensive set of collinear magnetic configurations.
Such configurations, as shown schematically in Fig.~\ref{MagConfigs}, include higher energy configurations besides the FiM ground state. 
These configurations are obtained by spin flipping of single magnetic units (M- and NIT-spin flips), or of dimer M-NIT units
in the  $\alpha$ and $\beta$ bonding (\emph{i.e.} frustrating only interactions of the $\alpha$- or $\beta$-type, see Fig.~\ref{MagConfigs}), 
and include the FM magnetic configuration as well.
Note that achieving the convergence for these selected magnetic configurations is not a trivial task,
as the free-electron on NITs tends to delocalize, hence a fine tuning of proper magnetic moments initializations and constraints was needed.   
The energetics of this set of magnetic configurations for CoPhOMe and MnPhOMe is reported in Tab.~\ref{EnergeticsMagConfig}.
These results pertain to the set $a$ of (U,J) parameters, but consistent values have been obtained also for sets $b$ and $c$ (not shown). 
For CoPhOMe, the FM state, with a magnetic moment of 12 ${\mu_{B}}$/cell, is  572 meV higher in energy respect to the FiM magnetic ground state.
Consistently, the energy cost of single Co- and NIT-spin flips is similar (184 meV). 
The most interesting result concerns the ($\alpha$ or $\beta$) dimer spin flips:  
there is a remarkable energy difference between the two configurations, with the $\alpha$-bond spin flip favored by 75 meV with respect to the $\beta$ one.   
Moreover, the (Co-NIT)$_{\alpha}$-spin-flip turns out to be by far the most favorable static magnetic excitation.
Moving to the Mn-chain, the FM state has a magnetic moment of 18 ${\mu_{B}}$/cell, and is 728 meV higher than the FiM ground state.  
The energy cost of a Mn-spin flip is 216 meV, and similarly to the Co-case, the (Mn-NIT)$_{\alpha}$-spin-flip is the favored magnetic excitation,
while the $\beta$ one is higher in energy by 40 meV.

An energy-mapping analysis within the broken-symmetry approach\cite{whangbo2013} was carried out to derive the exchange coupling constants, 
whereby the total energies obtained by first-principles calculations were then mapped onto a Heisenberg spin Hamiltonian:  
$\mathcal{H}_{\text{ex}} = \mathcal{H}_{\text{M-NIT}} + \mathcal{H}_{\text{NNN}}$ comprising nearest-neighbors (M-NIT) and next-nearest-neighbors (NNN) contributions:   
\begin{subequations} \label{equationa}
\begin{align}
\mathcal{H}_{\text{M-NIT}}=& -\sum_{r} \left[ 
J_\alpha\, \mathbf{S}_{2r}\cdot \mathbf{s}_{2r+1} + J_\beta\, \mathbf{S}_{2r}\cdot \mathbf{s}_{2r-1} \right]  \label{equationb}	\\
\begin{split}
\mathcal{H}_{\text{NNN}}=& -\sum_{r} \left[ 
J_\text{M-M}\, \mathbf{S}_{2r}\cdot \mathbf{S}_{2r+2} \right. \\
&\quad\left.+ J_\text{NIT-NIT}\, \mathbf{s}_{2r-1}\cdot \mathbf{s}_{2r+1} \right]  \label{equationc}	\,
\end{split}
\end{align}
\end{subequations}
with $\mathbf{S}_{2r}$ (even sites) indicating the spins of the metal ions and $\mathbf{s}_{2r+1}$ (odd sites) those of NIT radicals, 
where spins are assumed with unitary modulus in order to directly compare the strength of the exchange couplings in the two chains. 
The index $r$ labels different metal-NIT pairs in each chain. 
Equation~(\ref{equationb}) describes the coupling between the M-ion and the NIT-radical sublattices, the $\alpha$ and $\beta$ bondings contributing with a different coupling constant, 
whereas Eq.~(\ref{equationc}) takes into account both the M-M and NIT-NIT exchange interactions. 

\begin{table}
\caption{\label{EnergeticsMagConfig} Computed relative energies of the collinear magnetic configurations for CoPhOMe and MnPhOMe,
comprising the FiM, the M- and NIT-spin flips, the dimer (M-NIT)-spin flips, and the FM.
Calculations performed with U$_{M}$=5 eV (set $a$), energy values are in meV.}
\begin{ruledtabular}
\begin{tabular}{lll}
Magnetic config.		&	CoPhOMe		& MnPhOMe	\\ \hline
FiM					&	0.0				& 0.0			\\
M-spin-flip			&	184.1			& 216.0		\\ 
NIT-spin-flip			&	184.5			& 216.6		\\	
(M-NIT)$_{\alpha}$-spin-flip &	139.3			& 173.4		\\ 	
(M-NIT)$_{\beta}$-spin-flip   &   214.4 			& 213.7		\\ 	
FM 					&	572.2			& 727.7		\\ 
\end{tabular}
\end{ruledtabular}
\end{table}

\begin{table}
\caption{\label{MagInteractions} 
Exchange-coupling constants in CoPhOMe and MnPhOMe, as derived by Eq.~(\ref{equationa})
and with the convention of normalized spins in magnitude, computed for the three sets of (U,J) correlation parameters. Values are in meV.} 
\begin{ruledtabular}
\begin{tabular}{c c c c c}
CoPhOMe	& $J_{\alpha}$	& $J_{\beta}$ 	& $J_{\text{Co-Co}}$ & $J_{\text{NIT-NIT}}$ \\ \hline 
$a$			&	$-34.8$	&	$-53.6$	& $-1.7$			&	$-1.6$			     \\			
$b$			&	$-31.3$	&	$-45.9$	& $-1.3$			& 	$-1.4$			     \\					
$c$			&	$-23.8$	&	$-34.7$	& $-0.8$			&	$-0.4$			     \\	\hline \hline	
MnPhOMe	& $J_{\alpha}$	& $J_{\beta}$ & $J_{\text{Mn-Mn}}$ & $J_{\text{NIT-NIT}}$    \\ \hline	 
$a$			&	$-43.3$	&	$-53.4$	& $-6.6$			&	$-6.5$			     \\				 	
$b$			&	$-43.7$	&	$-53.9$	& $-6.7$			&	$-6.4$			     \\		
$c$			&	$-36.5$	&	$-44.3$	& $-5.0$			&	$-4.3$			     \\
\end{tabular}
\end{ruledtabular}
\end{table}
 
The estimated  exchange interactions are shown in Tab.~\ref{MagInteractions}.
Analyzing the CoPhOMe (set $a$), we first notice the large difference in the Co-NIT exchange coupling for the two bondings, $\beta$ being stronger by about 19 meV. 
Note also the presence of longer-range NNN interactions between Co's and NIT's spins of AF-type (of equivalent strength, about 1.6 meV).
Moreover, by including a finite J in the GGA+U Lichtenstein approach on Co-sites (set $b$), and by increasing further the correlation on O and N atoms (set $c$),
the exchange-coupling constants are subject to a reduction, as expected. 
We note that a uniform reduction for the NN-interactions is found, expressed by a constant ratio $J_{\alpha}/J_{\beta}$ $\sim$0.7 for the three (U,J)-sets here considered. However, this is not the case for the NNN-interactions, whereby the $J_{\text{NIT-NIT}}$ is more significantly reduced by the inclusion of U on the O- and N- atoms
than the $J_{\text{Co-Co}}$. This can be understood considering that the exchange coupling $J_{\text{NIT-NIT}}$ involves two NITs and an intermediate Co-atom,
and it is therefore more affected by an increased localization onto the NITs atoms.
The value of the Co-NIT exchange coupling constant here calculated is larger than the experimental estimation obtained by magnetic susceptibility measurements\cite{caneschi2001}.
However, it is important to emphasize that the experimental estimates of the exchange couplings strongly depend on the spin Hamiltonian that is assumed to compute the magnetic susceptibility.
This is crucial for CoPhOMe, for which the assumptions of a pure Ising model turns out to be inadequate, as it will be shown in Sec.~\ref{Thermodynamics},
where the thermodynamic properties of the system are derived by using a more sophisticated effective spin-Hamiltonian based on density functional results.  
Note moreover that these exchange energy values are consistent with \emph{ab initio} calculations within Quantum Chemistry approach carried out on a simplified cluster model  
composed of a radical Co-NIT pair (in the $\alpha$-bonding)\cite{Bogani_NatMat13}. 
The authors obtained a value $J_{\alpha}$ $\sim$ -163.5 cm$^{-1}$ (-20.3 meV), evaluated from the energy gap between the two lower spin-states,
\emph{i.e.} the triplet (S=1) and the quintet (S=2) spin-states corresponding to the AF and FM coupling of the Co-NIT spin-pairs, respectively.

Moving to MnPhOMe, the computed exchange interactions are AF-type, both Mn-NIT and NNN ones, and are considerably stronger than those found in the Co-chain.
Particularly, the $\alpha$ bondings are stronger than the corresponding Co ones by about 10 meV. Importantly, the NNN interactions are here sizable (about 6.5 meV).
Similarly to CoPhOMe, the magnetic exchange at the $\alpha$ bond is weaker than at the $\beta$ one, with a consistent ratio $J_{\alpha}/J_{\beta}\sim 0.8$ for all the considered sets.  
Experimentally, among the Mn-NIT chain complexes, MnPhOMe has been characterized as the compound with the largest value of antiferromagnetic coupling constant\cite{caneschi1991}.
By using a NN Heisenberg model to reproduce the experimental susceptibility, an exchange coupling constant $J_{\text{Mn-NIT}} = -344$ cm$^{-1}$ ($-42.6$ meV)
was estimated. 
Due to the negligible anisotropy of Mn(II)-ions, the \textit{g} factor of both Mn and NIT spins was fixed to 2.0 value for the fitting. 
Our results are in good agreement with this experimental value, and fully justify the choice of an isotropic Heisenberg model (see below).

Some considerations about the calculated exchange coupling constants are here given.
(\emph{i}) The magnetic building units in both compounds are chemically rather similar, 
\emph{i.e.} Co or Mn ions coupled via direct type magnetic exchange to nitronyl-nitroxide radicals; 
the main difference consisting in an enhanced exchange pathway for the Mn-compound, due to the half-filled \emph{d}-shell electronic configuration. 
Therefore, our results of exchange coupling constants of strength of the same order of magnitude for both complexes,
yet with a stronger coupling for the MnPhOMe, are fully consistent based on these qualitative arguments.
(\emph{ii}) The strength of the exchange coupling strongly depends on the structural parameters of the magnetic constituents. 
As a matter of fact, a change in structural parameters such as bond lengths and angles 
will modify the degree of overlap of the local magnetic orbitals contributing to the exchange pathways, 
and this will impact on the resulting strength of the exchange coupling.  
Recently, the magneto-structural correlations between Mn(II) ion and NITs in \emph{trans-} and \emph{cis}-coordinated model complexes 
were investigated by broken-symmetry DFT approach\cite{wei2005}.
The authors found that the exchange coupling constant $J_\text{Mn-NIT}$ increases linearly by decreasing the bond distance Mn-O (O of the NIT) in these complexes.
If we consider solely the effect of the Mn-O bond length variation going from $\alpha$-bond (2.14 {\AA}) to the $\beta$-one (2.12 {\AA}) in MnPhOMe,
from their results we can roughly estimate an increase $\Delta J_{\text{Mn-NIT}}$ of $\sim$ 6\%. 
Other structural parameters are involved as well, and that may account for the variation of $\sim$ 20\% for the two bondings in MnPhOMe, as indicated by our calculations.
In this regard, bond angles variations are even more important, as they can induce variations in the exchange coupling constants up to 40 -- 50\% \cite{wei2005}.
(\emph{iii}) Finally, it is worth emphasizing that the method adopted -- by considering the whole experimental structure within a periodic boundary approach 
and examining various magnetic configurations -- allowed us to enrich the picture of the magnetic interactions at play: 
quantities which are difficult to access experimentally and neglected within a cluster-model approach.
Notably, although metal-ions are far apart (the intra-chain Co-Co and Mn-Mn distances are 7.815 {\AA} and 7.878 {\AA}, respectively),
we do appreciate non-negligible NNN-exchange interactions, which overall may be particularly relevant for the specific case of MnPhOMe.
Interestingly, in analogous chains comprising lanthanide ions and NIT radicals NNN interactions are dominating\cite{Bartolome1996}.

\subsubsection{\label{Magnetic anisotropy}Magnetic anisotropy} 
The anisotropy of the Co-compound is ultimately of magneto-crystalline type,
namely single-ion anisotropy deriving from the $d^7$ electronic configuration in the octahedral field, with a hole in the $t_{2g}$ orbitals.
This electronic structure yields a sizable orbital magnetic moment on the Co-sites, 
and this leads to strong anisotropy due to the large first-order spin-orbit coupling. 
The ion Co$^{2+}$ is therefore susceptible to the anisotropy of the chemical environment, 
whereas this is not the case for the Mn$^{2+}$, which is in $d^5$ configuration and with negligible orbital angular momentum. 
This leads to the different magnetic behavior of the two systems.
Due to the low symmetry of the metal-sites local chemical environments and their distortions from octahedral symmetry,
a direct calculation of the anisotropy for the real chemical systems is needed.
The characterization of the local magnetic anisotropy of the M-sites in the chains turns out to be unfeasible
due to the impossibility to single out the single-ion anisotropy from the exchange interactions with the confining radicals.
To this purpose, we have hence resorted to the molecular compound Co(hfac)$_{2}$-(NITPhOMe)$_{2}$, hereafter called CoNIT$_{2}$,
which is essentially the mononuclear variant of the chain, containing the octahedrally coordinated Co(II) ions with two hfac molecules
and two NITPhOMe ligands. 
The latter are coordinated in \textit{cis} configuration to the central Co-ion through a single ON group\cite{caneschi2002}.

Since the metal ion has essentially the same local coordination environment in the mononuclear compound and in the chain,
the two complexes are expected to be characterized by the same local anisotropy properties. 
The Co single-ion anisotropy in the compound CoNIT$_{2}$ was evaluated by performing magnetic calculations (in the non-collinear formalism and inclusive of SOC) 
for the spin-trimer magnetic structure, \emph{i.e.} with the spins of the  NITs collinear among them and AF-coupled to the spin of the intermediate Co-atom.
In this way, the two Co-NIT exchange interactions are fully satisfied and the energy variations obtained by rotating the orientation of the (total) spin-trimer axis
can be safely attributed entirely to the Co single-ion anisotropy, since the spins of the attached radicals are isotropic.      
The experimental structure of the CoNIT$_{2}$, which crystallizes in the triclinic $P\bar{1}$ space group\cite{caneschi2002}, was considered for the calculations.
The Magnetic Anisotropy Energy (MAE) was characterized by total energy calculations 
with the axis of the total spin-trimer spanning the three planes of a Cartesian system XYZ (in steps of 15$^{\circ}$).

Regarding the Mn-compound, the experimental structure is not available in literature.
Therefore, we used the structure of the CoNIT$_{2}$ and replaced the central Co atom by Mn.  
The Mn-O bond lengths in the octahedral complex are shorter in this model system compared to the experimental structure of the MnNIT$_{2}$ compound,
and this is expected to slightly overestimate the magnetic exchange between the Mn-ion and radicals. 
However, being aware of this limitation of the model, we resort here to the molecular cluster solely to inspect the single-ion anisotropy,   
and we believe this should not affect significantly the results for this specific purpose.

As an additional methodological remark, we found that the magnetic anisotropy critically depends on the parameter J in the GGA+U formalism, 
while the role of the parameter U is negligible. 
Specifically, whereas the anisotropy profile is overall consistent, with location of minima and maxima in the energy landscape being unaltered,
the strength of the anisotropy enhances with increasing J values.
Based on that, the results here presented are relative to set $c$, with M-ions subject to a sizable J value of 1.5 eV.
As a matter of fact, calculations with J=0, \emph{i.e.} set $a$, show a reduced anisotropy profile.
It is worth noting that non-collinear calculations were performed by achieving full self-consistency after the inclusion of SOC-term with a more accurate energy threshold,
as we realized that standard magnetic calculations with non-selfconsistent SOC cycle yield anisotropy properties largely quenched.
The non-collinear magnetic configurations were generated by adopting the constraining technique of the local magnetic moments orientation,
as implemented in the VASP code. 
This approach is essentially based on the inclusion of an energy penalty term to the total energy that drives the local moments to the desired directions.
We set the energy threshold to 10$^{-7}$ eV and, after proper initialization of the magnetic moments along defined spatial directions of the total spin,
increased stepwise the weight of the penalty term, until convergence of the non-collinear constrained SOC calculations was achieved.

\begin{figure}
\includegraphics[width=1\linewidth]{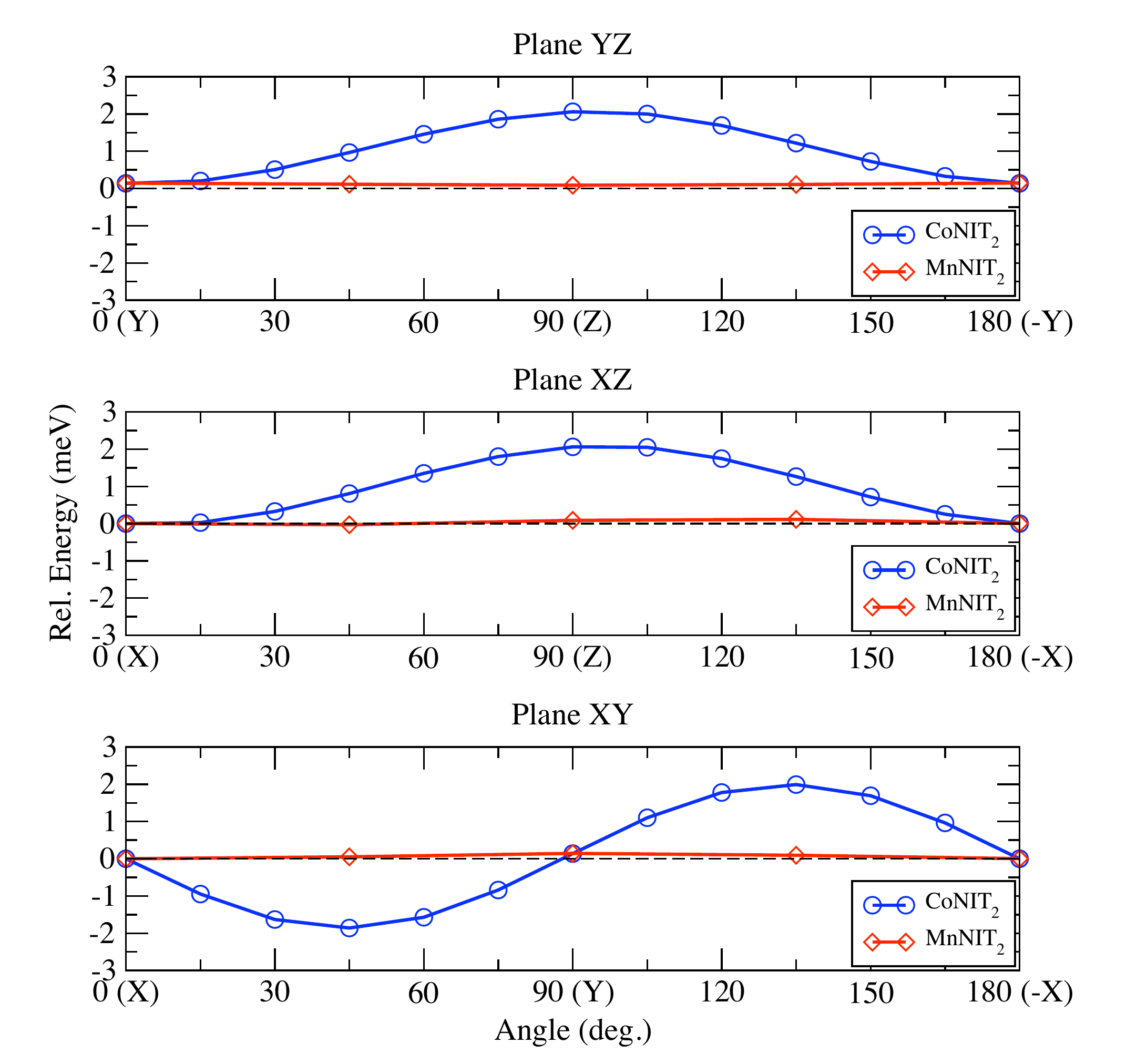}
\caption{\label{MAE}  Magnetic Anisotropy Energy profiles for Co- and Mn-based molecular compounds CoNIT$_{2}$ and  MnNIT$_{2}$,
as the orientation of the total spin is varied along three planes of a Cartesian system XYZ.
The axes (X,Y,Z) of the reference Cartesian system are related to the crystallographic directions ($a, b, c$) of the triclinic lattice as follows:
$a \equiv$ X; $b$ lies on the XY-plane at 64.51$^{\circ}$ with respect to X-axis, the $c$-axis is tilted of 11.97 $^{\circ}$ with respect to the Z-axis,
and with projection on the XY-plane lying on the first quadrant. Energies are relative to the value on the X-axis direction.}
\end{figure}

MAE profiles for both compounds are shown in Fig.~\ref{MAE}. 
It is evident that, while the Mn-compound is nearly isotropic, with the largest energy variation of about 0.15 meV,
the CoNIT$_{2}$ shows a sizable magnetic anisotropy with the largest energy barrier $\sim$ 3.85 meV (46K).
Additionally, upon closer inspection, the anisotropy on each Co-ion turns out to be uniaxial. 
This is evidenced by the markedly deeper energy barrier for magnetization reversal experienced by the spin-trimer in the XY-plane  
compared to the shallower barriers (saddle points) along the other two sampled planes.        
These results are in good agreement with the experimental characterization of the MnPhOMe as an isotropic Heisenberg-like FiM spin chain,  
as well as with the description of the CoPhOMe as a non-collinear FiM spin helix, deriving from uniaxial character of Co-octahedral complexes.
In this second spin chain, the easy axis of each Co is not collinear with respect to the ones of neighboring Co-atoms but follows the three-fold periodicity of the crystal.  
By mapping the estimated local easy-axis of the Co ion in the molecular compound onto the crystal frame of the chain, 
the tilt angle formed by the Co spins relative to the chain axis \emph{c} (polar angle $\theta$) can be estimated, yielding $\theta=$ 50 $\textendash$ 55$^{\circ}$. 
This value is in reasonable agreement with the experimental estimate given in Ref.~\onlinecite{CaneschiEPL2002} (see below).     
However, a direct comparison of the MAE obtained by DFT-calculations with the EPR-characterization of the anisotropy\cite{caneschi2002} in the compound CoNIT$_{2}$
shows some inconsistencies regarding the absolute orientation of the anisotropy easy-axis. 
This aspect requires further investigation and will be the subject of future work.
Notwithstanding, it is worth emphasizing that this discrepancy between theory and experiments about the absolute easy-axis direction in the mononuclear compound 
will impact  -- once mapped onto the chain crystal frame -- only on the azimuthal angle $\phi$ (\emph{i.e.} the orientation on the \emph{ab} plane),
which is overall not relevant due to the chain trigonal symmetry, while the polar angle $\theta$ is well reproduced. 

Additionally, the anisotropy was assessed directly in the crystal frame of the chain compound. 
A model system was used for this purpose in which two Co atoms were replaced by Zn's in the unit cell of CoPhOMe.
In this way, the diamagnetic Zn(II) ions turn off the magnetic interactions with the neighboring radicals, 
and the magnetic complex is hence simplified to a spin-trimer comprising the Co-site AF-coupled to two lateral NITs, though in the helical structure.
By rotating the axis of the total spin of the complex along defined directions, the anisotropy of the local Co-octahedral environment was probed. 
The set of directions included the Co-O axes and bisectors of the octahedral local frame. 
Due to the demanding computational task required, the SOC term was included in a non-self-consistent way. 
Although the comparison remains qualitative since non-self-consistent calculations do not account for the whole anisotropy strength, 
the results are in accord with those performed on the mononuclear-complex, namely easy-axis as well as energy valleys/rises of the MAE profile in the two local frames coincide. 

Static measurements of the magnetization in CoPhOMe show that the magnetization relaxes slowly below a temperature of 6 K when an external magnetic field is applied parallel to the trigonal axis, while no (dynamic) hysteresis is observed for magnetic fields applied on the trigonal plane\cite{caneschi2001}. 
This observation can be ascribed to non-collinearity among anisotropy axes, as it will be discussed in Sec.~\ref{Energy_scale}  
%

\section{\label{Thermodynamics}Finite temperature properties}  
The thermodynamics of spin chains is dictated by the temperature dependence of the correlation length $\xi$, i.e., the characteristic scale of decay of pair-spin correlations. 
For infinite chains, this intrinsic length scale is proportional to the product of the static susceptibility ($\chi$) measured in zero magnetic field and temperature: $\chi\,T\sim \xi$. 
The last relation is usually employed to extract information about the spin-Hamiltonian parameters directly from experimental susceptibility data. 
In the absence of anisotropy (as in the case of MnPhOMe) the correlation length is expected to diverge like $J_{\rm ex}/T$ with decreasing temperature ($J_{\rm ex}$ being some effective
exchange constant describing the coupling between nearest neighbors). 
In spin chains with uniaxial anisotropy, one expects $\xi\sim {\rm e}^{\Delta_{\xi}/ T}$, where the characteristic energy scale $\Delta_{\xi}$ 
may have a residual dependence on $T$ arising from renormalization due to spin waves~\cite{Sangiorgio2014} ($k_B=1$ will be assumed henceforth). 
Such a mechanism is not effective when elementary excitations are sharp (Ising) DWs~\cite{Billoni2011}, which happens when the uniaxial anisotropy is comparable to  
the exchange interaction or larger.  
In this case, $\Delta_{\xi}=2J_{\rm ex}$ throughout the whole range of temperatures where $\xi$ exceeds some lattice units 
(assuming a unitary spin modulus). 
When the exchange energy is larger than the anisotropy energy, elementary excitations are broad DWs and the relationship between $\Delta_{\xi}$ -- accessible in experiments -- 
and spin-Hamiltonian parameters is not straightforward anymore. 
The anisotropy energy computed for the CoNIT$_{2}$ and the estimates of $J_{\alpha}$ and $J_{\beta}$ obtained for CoPhOMe suggest that in this spin chain
elementary excitations should actually consist of broad DWs.  
The scenario is further complicated in CoPhOMe by the non-collinearity among local anisotropy axes.   
In the following, the question of how the energy barrier $\Delta_{\xi}$ is affected by the degree of non-collinearity will be addressed. 

\subsection{Model and results}  
To model the thermodynamics of CoPhOMe we propose the Hamiltonian  
\begin{equation}
\label{Ham_TD}
\begin{split}
\mathcal{H}&= -\sum_{p,r} \left[ 
J_{\rm ex}\,\mathbf{S}_{p,2r}\cdot(\hat{\boldsymbol{\sigma}}_{p,2r+1} + \hat{\boldsymbol{\sigma}}_{p,2r-1}) \right. \\
&+\left.  \mathbf{S}_{p,2r}\widetilde{D}_{2r}\mathbf{S}_{p,2r} +
\mu_B \left(g \mathbf{B}\cdot\hat{\boldsymbol{\sigma}}_{p,2r+1} +  G\,\mathbf{B}\cdot\mathbf{S}_{p,2r}\right)
\right]
\end{split}
\end{equation}
where  Co spin operators are replaced by classical vectors  $|\mathbf{S}_{p,2r}|=1$  and $\hat{\boldsymbol{\sigma}}_{p,2r+1}$ are Pauli quantum operators, representing NIT spins. 
A uniaxial anisotropy was assumed for each Co in the corresponding local frame, namely $D_x=D_y=0$ and $D_z=46$ K (obtained from DFT calculations on CoNIT$_{2}$). 
Each of these anisotropy tensors is reported to the crystal frame by a different rotation parameterized by a standard  matrix    
$\mathcal{R}(\varphi,\theta,\psi)$, where $\varphi,\theta,\psi$ are Euler angles. 
More explicitly $\widetilde{D}_{2r}=\mathcal{R}(\varphi,\theta,\psi)\,{\rm diag}(D_x,D_y,D_z) \mathcal{R}^{T}(\varphi,\theta,\psi)$,  
where ${\rm diag}(D_x,D_y,D_z)$ indicates the 3$\times$3 diagonal matrix with eigenvalues $D_x$, $D_y$ and $D_z$.  
The crystal symmetry imposes  $\varphi=2\pi\,r/3$ with $r=1,2,3$ so that the index $r$ labels the atoms inside the cell. $p$ is the cell index and ideally extends to infinity.  
Due to our choice of vanishing $D_x$ and $D_y$, the tensor $\widetilde{D}_{2r}$ is independent of the $\psi$ angle. 
Therefore, the degree of non-collinearity is \textit{uniquely} parameterized by $\theta$: For $\theta=0^{\circ}$ anisotropies are collinear and 
point along the chain axis $c$; while for $\theta=90^{\circ}$ local anisotropy axes lie on the \emph{ab} plane, with neighboring axes forming an angle of 120$^{\circ}$.  
The coupling between Co ions and the magnetic field should properly be described by a Land\'e tensor. However, 
since the Zeeman term does not affect the energy barrier $\Delta_{\xi}$, a scalar Land\'e factor was assumed for both Co atoms and NITs, with $G=2\times 3/2=3$ and $g=2\times 1/2=1$ (including the spin modulus).  
Moreover, only the exchange coupling between each Co and the neighboring radicals was considered and set to $J_{\rm ex}=-500$ K. 
This value is intermediate between $J_\alpha=-404$ K and $J_\beta=-622$ K, obtained from the  set $a$ of DFT calculations.  
In Appendix~\ref{Continuum-limit} it is shown that distinguishing between 
$J_\alpha$ and $J_\beta$ does not alter the thermodynamic properties of this model in the temperature range of interest and for this choice of parameters.     
The mapping of the model with $J_\alpha \ne J_\beta$ into a model with a unique exchange coupling  $J_{\rm ex}$ relies on assuming broad DWs. 
Therefore, for values of $D_z$ consistent with the formation of sharp DWs this equivalence is not guaranteed.  

Equilibrium properties at finite temperature can be deduced from the knowledge of the partition function $\mathcal{Z}$, which is related to the free energy $F$ by the fundamental equation 
$F= -T \ln\left(\mathcal{Z}\right)$. The magnetization and the susceptibility are given by the first and the second derivative of $F$ with respect to the applied field. 
These quantities have been computed for the infinite chain, i.e. $N_p\rightarrow\infty$, with the transfer-matrix technique. 
Details about this method, including the treatment of quantum operators  $\hat{\boldsymbol{\sigma}}_{p,2r+1} $, are given in Appendix~\ref{Transfer-Matrix}. 
%
\begin{figure}[ht]
\begin{center}
\includegraphics[width=1\linewidth]{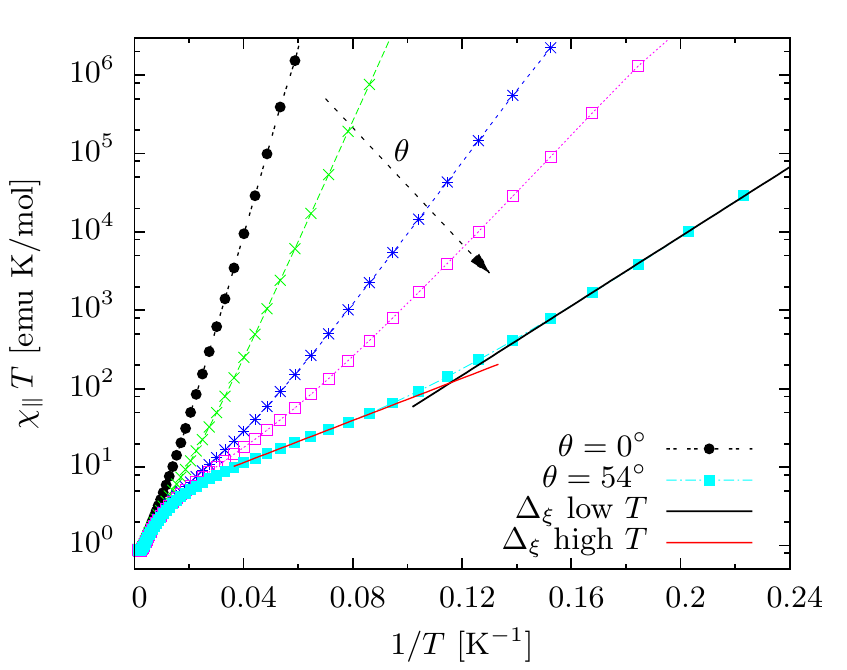}  
\end{center}
\caption{Color online. Susceptibility along the chain axis $\chi_\parallel=\partial M_\parallel/\partial B_\parallel$  computed in $B=0$ for 
$\theta=0^\circ$ (dots), $40^\circ$ (crosses), $50^\circ$ (stars), $52^\circ$ (empty squares), $54^\circ$ (full squares).  
The arrow indicates the direction of increasing $\theta$. Solid lines represent the linear fittings of 
low- and high-temperature barriers: $\Delta^{\rm low}_\xi$ (black line) and $\Delta^{\rm high}_\xi$ (red line). \label{Fig_chi_par} }
\end{figure} 

\begin{figure}[ht] 
\begin{center}
\includegraphics[width=1\linewidth]{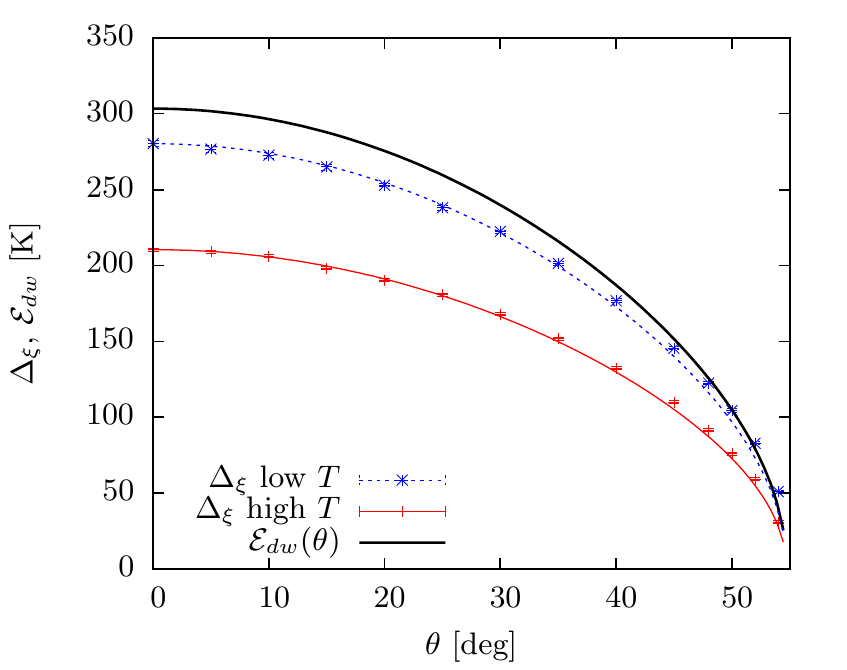} 
\caption{Color online. Energy barriers $\Delta^{\rm low}_\xi$ (blue stars) and $\Delta^{\rm high}_\xi$ (red crosses) extracted from the susceptibility curves at low and intermediate temperatures, as described in the text, are plotted with symbols as a function of $\theta$. 
Error bars are determined by the uncertainty on the fitted slopes.   
The upper, black line represents the DW energy $\mathcal{E}_{\rm dw}(\theta)$; the dashed blue and the red lines represent the same quantity rescaled to the value of $\Delta_\xi$ obtained for $\theta=0$ at low and high temperature, respectively (see Eq.~\eqref{Delta_xi_CoPhOMe}).  
\label{Fig_Delta_xi}}
\end{center}
\end{figure}

We focused on the susceptibility $\chi_\parallel=\partial M_\parallel/\partial B_\parallel$ along the chain axis \emph{c} computed in zero field for different values of the angle $\theta$. 
In Fig.~\ref{Fig_chi_par} these results are plotted as $\chi_\parallel\,T$ versus $1/T$ in a log-linear scale in order to highlight the exponential divergence of the correlation length $\sim{\rm e}^{\Delta_{\xi}/ T}$ and 
possible deviations from it. 
For each value of $\theta$, two ``linear regimes'' -- corresponding to two values of $\Delta_\xi$ --  can be identified at low and intermediate $T$. 
For a given $\theta$, the value of $\Delta_\xi$ is systematically larger at low $T$ than at hight $T$.
Black and red lines are drawn on top of the $\chi_\parallel\,T$ curve computed for $\theta=54^\circ$ to exemplify the two linear fittings.   
Enlarging the horizontal scale two distinct slopes can be highlighted for smaller $\theta$ angles as well. 
The two energy barriers  corresponding to each $\theta$  are plotted with symbols in Fig.~\ref{Fig_Delta_xi}, 
blue stars for the low-temperature  $\Delta^{\rm low}_\xi$ and red crosses for the high-temperature $\Delta^{\rm high}_\xi$.  
As non-collinearity among anisotropy axes is increased, by increasing $\theta$, both high- and low-temperature 
values of $\Delta_\xi$ decrease: the former passes from 210.4 K in the collinear case ($\theta=0^\circ$) to 30.9 K for $\theta=54^{\circ}$; the latter ranges from 280.2 K to 50.9 K when $\theta$ is varied in the same interval.   
The existence of two energy barriers $\Delta^{\rm low}_\xi$ and $\Delta^{\rm high}_\xi$ is not due to non-collinearity. In fact, this phenomenon occurs also in ferromagnetic Heisenberg chains with collinear anisotropy axes for ratios of $D_z$ to the exchange energy compatible with the formation of broad DWs~\cite{Billoni2011,Sangiorgio2014,gatteschi2013}.   
As anticipated, this is in contrast to what happens in the limit of sharp DWs, where $\Delta_\xi$ takes a single value, right equal to the DW energy. 
Moreover, in Heisenberg chains with broad DWs and collinear anisotropy $\Delta_\xi$ is always smaller than the DW energy. 
At low temperature this is due to the interplay between DW excitations and spin waves.  
More precisely, for each DW that is added to the system, 
two Goldstone modes appear in the spin-wave spectrum\cite{Yan_12} that are associated with translational invariance of the DW center and with the degeneracy with respect to the azimuthal angle (e.g., Bloch and N\'eel DW have the same energy if $D_x=D_y$)\cite{Sangiorgio2014}. Besides this, the presence of DWs also modifies the density of states of spin waves with respect to the case in which the last ones are superimposed to a uniform spin profile. All this results in an entropic contribution that affects the correlation length and makes the low-temperature $\Delta^{\rm low}_\xi$ smaller than the DW energy\cite{Fogedby84JPCSSP,HB_Braun_PRB94}. Still in collinear chains, the further suppression of $\Delta_\xi$ occurring at intermediate temperatures has been justified in the framework of Polyakov renormalization of spin Hamiltonian parameters\cite{Billoni2011,Sangiorgio2014}. 

By analogy with the collinear case, we may infer that the suppression of $\Delta_\xi$ 
with increasing temperature observed in our \textit{non-collinear} model -- at fixed $\theta$ -- also emerges from the interplay between spin-wave and DW excitations.      
In fact, with some additional hypotheses based on $J_{\rm ex}\gg D_z$ (see Appendix~\ref{Continuum-limit}) and setting $\mathbf{B}=0$, Hamiltonian~\eqref{Ham_TD} can be mapped into the following functional 
\begin{equation}
\label{Ham_continuum_2}
\mathcal{H}= \int \left[ \frac{|J_{\rm ex}|}{4}\left(\partial_z \mathbf{S}\right)^2 -  \frac{1}{2} D_z \left(3\cos^2\theta-1\right) \left(S^z\right)^2 \right] dz \,.
\end{equation} 
This functional is formally the same as  the  continuum version of the Heisenberg Hamiltonian with \textit{collinear} anisotropy axes 
discussed before.  
Note that the penalization due to the misalignment of neighboring spins is halved with respect to the ferromagnetic Heisenberg  chain ($|J_{\rm ex}|/4$ instead of $J_{\rm ex}/2$), because the coupling between 
two consecutive classical (Co) spins is mediated by the quantum spin of the interposed NIT, and the uniaxial anisotropy is replaced 
by the $\theta$-dependent term $D_z \left(3\cos^2\theta-1\right) /2$. 
Thanks to this mapping, the analytic expression giving the cost to create a DW in the Heisenberg chain with collinear anisotropies\cite{Enz} can be used to get 
$\mathcal{E}_{\rm dw}(\theta)=\sqrt{2 |J_{\rm ex}|D_z \left(3 \cos^2\theta -1\right)}$. 
Remarkably, the effective anisotropy (and $\mathcal{E}_{dw}(\theta)$ with that) vanishes at the magic angle $\theta=54.7^\circ$, 
above which both the magnetization and the susceptibility become larger on the \emph{ab} plane than along the \emph{c} axis (parallel to $z$). $\mathcal{E}_{\rm dw}(\theta)$ is plotted as a solid black line in Fig.~\ref{Fig_Delta_xi}. 
In collinear chains both high- and low-temperature $\Delta_\xi$ are smaller than $\mathcal{E}_{\rm dw}$ but proportional to it\cite{Billoni2011}. According to Fig.~\ref{Fig_Delta_xi},  
this trend seems to be maintained in CoPhOMe, in spite of its non-collinear structure. To better verify this trend we rescaled $\mathcal{E}_{\rm dw}(\theta)$ to the  high- and low-temperature
values taken by $\Delta_\xi$ in the collinear case ($\theta=0^\circ$). The corresponding curves are plotted with lines of the same colors as the fitted barriers $\Delta_\xi$ (symbols). 
The agreement is quantitative within 5\% up to $\theta=45^\circ$, while larger deviations appear when $\theta$ approaches the magic angle. 
For $\theta<45^\circ$ we can, therefore, express $\Delta_\xi$ in terms of 
the spin-Hamiltonian parameters and the $\theta$ angle as follows: 
\begin{equation}
\label{Delta_xi_CoPhOMe}
\begin{split}
&\Delta^{\rm high}_\xi\simeq 0.69\, \mathcal{E}_{\rm dw}(\theta)=0.69 \sqrt{2 |J_{\rm ex}|D_z \left(3 \cos^2\theta -1\right)}\\ 
&\Delta^{\rm low}_\xi\simeq 0.92\, \mathcal{E}_{\rm dw}(\theta)=0.92 \sqrt{2 |J_{\rm ex}|D_z \left(3 \cos^2\theta -1\right)} 
\end{split} 
\end{equation} 
where the superscripts stand for high- and low-$T$, respectively. 

For the sake of completeness,  we mention that Hamiltonian~\eqref{Ham_TD} is suitable to model MnPhOMe as well, provided that the anisotropy term is removed. 
In zero applied field the relevant observables can be computed analytically\cite{Seiden}.

\subsection{\label{Energy_scale}Comparison with experiments}
The angle that anisotropy axes form with the $c$ axis in CoPhOMe is estimated to be $\theta \simeq 50^\circ$ or larger by DFT calculations. 
A similar estimate was already reported in the early works on this compound based on the (small) difference between the saturation magnetization measured along the chain axis and on the \emph{ab} plane\cite{CaneschiEPL2002,VindigniPRB04}.   
For this value of $\theta$ and the computed parameters (set $a$) the DW energy is 
$\mathcal{E}_{\rm dw}(50^\circ)=105$ K.  
%
In the following we will focus on the high-temperature $\Delta^{\rm high}_\xi$, which is the only barrier accessible experimentally. In fact, finite-size effects  
-- induced by the presence of naturally occurring defects\cite{BoganiPRL2004,VindigniAPL05} -- and slow dynamics prevent the correlation length, and $\chi$ consequently, from diverging   
indefinitely as temperature is lowered.  
The numerical transfer-matrix calculation yields $\Delta^{\rm high}_\xi=76$ K, which falls in the correct range, even if a more quantitative comparison of this thermodynamic model with experiments would require a direct fitting of the susceptibility data. 
At present, it is important to stress that the observation of $\Delta_\xi$ of the order of hundred K does not necessarily imply that    
$|J_{\rm ex}|$ be of the order of 50 K, as the Ising model would prescribe: 
With the more realistic model Hamiltonian in Eq.~\eqref{Ham_TD} an estimate of $J_{\rm ex}=-500$ K obtained from \emph{ab-initio} calculations can be absolutely 
consistent with a much smaller $\Delta_\xi$.

When it was firstly synthesized, the novelty of CoPhOMe consisted in displaying slow relaxation of the magnetization. 
The relaxation time was observed to obey an Arrhenius law $\tau=\tau_0 {\rm e}^{\Delta_\tau/T}$ over ten decades, which was explained in the framework of the Glauber model\cite{CoulonPRB07,CoulonPRB04,VindigniAPL05,VindigniJPCM09,PiniPRB11}. 
Strictly speaking, the Glauber model is a kinetic version of the Ising model, which assumes a very large (ideally infinite) anisotropy. 
Even if in realistic SCMs the anisotropy is finite, this description has proven to be appropriate at the condition that elementary excitations are sharp DWs.  
In particular, at low temperature the relaxation process initiates by reversing a spin lying at one of the two edges of each open chain (see below), with an activation energy equal
to the DW energy and, therefore, to $\Delta_\xi$. The diffusion of a sharp DW requires an additional activation energy (usually named $\Delta_A$) at each step\cite{CoulonPRB04,Coulon06Springer}. 
Finally, in the large-anisotropy limit, one has  $\Delta_\tau=\Delta_\xi+\Delta_A$.  

The fact that DFT calculations yield $|J_{\rm ex}| \gg D_z$ calls for a revision of the mechanism behind slow relaxation and the relative energy scales, because if elementary excitations are \textit{broad} DWs the Ising and the Glauber models are not applicable. So far there is less quantitative agreement between theory and experiments on SCMs with broad DWs and the interpretation of available data is still controversial. 
However, theoretical arguments suggest that the diffusion of a broad DW does not proceed by thermal activation\cite{Billoni2011} (i.e. $\Delta_A=0$).  
The energy cost to nucleate a DW from one edge of an open spin chain is expected to be of the order of the DW energy at zero temperature $\mathcal{E}_{\rm dw}(\theta)$.  
A later experimental study on CoPhOMe -- in which the amount of non-magnetic impurities was varied in a controlled way -- confirmed that the 
relaxation precess indeed initiates at one edge of a finite segment of coupled spins\cite{BoganiPRL2004,VindigniAPL05}. 
%
Since in CoPhOMe $\theta$ is certainly larger than $45^\circ$, Eq.\eqref{Delta_xi_CoPhOMe} does not hold exactly. However, this equation can still help estimate the relation between 
$\Delta^{\rm high}_\xi$ and $\Delta_\tau$, the latter beeing identified here with the DW energy: 
one expects the measured $\Delta_\xi$ to be roughly 70\% of $\Delta_\tau$. 
From the value of $J_{\rm ex}$ used in Ref.~\onlinecite{caneschi2001} to fit the susceptibility data, we deduce that $\Delta_\xi\simeq 110$ K for that specific sample; 
in the same reference it is reported that $\Delta_\tau=154$ K, which yields a ratio between these two energy barriers in fair agreement with our expectation. 
In conclusion, this alternative paradigm based on broad DWs seems plausible. 
Further details about how the diffusion of a broad DW can be mapped into the Glauber model are given in Appendix~\ref{DW_diffusion}.

The Glauber model successfully justifies the observation of slow relaxation in CoPhOMe only for magnetic fields applied along the chain axes.  
As already mentioned, this originates from non-collinearity among anisotropy axes\cite{CaneschiEPL2002,VindigniJPCM09}.  
By virtue of Hamiltonian~\eqref{Ham_continuum_2}, the same phenomenon can be justified in terms of the behavior expected for Heisenberg chains with  collinear anisotropy.  
As far as $\theta$ is smaller than the magic angle, the effective anisotropy $D_z \left(3\cos^2\theta-1\right) /2$ is positive and the \emph{global} easy axis points along \emph{c}.   
The creation of a DW by thermal activation is thus required to reverse the magnetization along the \emph{c} axis ($\parallel z$). Conversely, rotating the magnetization on the \emph{ab} plane should not require any activation energy. For larger $\theta$, the spin chain becomes easy-plane, but still the magnetization reversal on the \emph{ab} plane should not be a thermally activated process because  in this case   
the system should behave similarly to the XY model. 

\section{\label{Summary}Conclusions and perspectives}  
In this work, we have addressed -- by means of density functional calculations -- the study of two archetypical organic magnetic chains, 
composed of M(hfac)$_{2}$ moieties bridged by NITPhOMe radicals, with M=Co, Mn. 
They are structurally very similar, but their magnetic behavior is remarkably different. 
CoPhOMe chain is characterized by SCM behavior, whereas MnPhOMe is described as an isotropic spin chain.
A detailed investigation of the magnetic, electronic, and anisotropy properties has been performed, enriching the understanding of these systems, 
with particular emphasis on the description at the atomic scale.
 
Consistently with experiments, we find that both chains have a ferrimagnetic ground state, 
as a result of the local AF-coupling between the metal-ion and the radical.   
The exchange interaction M-NIT is rather strong, of the order of 30 \textendash\ 50 meV for CoPhOMe, and even of greater strength for MnPhOMe. 
The latter is characterized also by sizable next-nearest neighbor exchange interactions (between Mn-Mn and NIT-NIT couples) of AF-type.
Importantly, we find that there are two types of M-NIT exchange couplings, of considerably different strength.
This finds a correspondence in their structural features, whereby the bridging NITs are characterized by two non-equivalent bondings with the confining metal ions. 
The difference in CoPhOMe is about 19 meV, while in MnPhOMe it is $\sim$ 10 meV .   
 
The electronic properties are pretty similar in the two chains, both systems behaving as small gap semiconductors.
The HOMO and LUMO states are essentially composed of the $p$-states of the spin-polarized O and N atoms of the NITs, 
while the occupied $d$-states of the transition metals are quite far from the gap, strongly hybridizing with the O states. 
The magnetic anisotropy has been characterized in the mononuclear variant  of the chains, the M-NIT$_{2}$ compound, which is a magnetic spin-trimer.
We find that the Mn-compound is almost isotropic, 
while the Co-compound is characterized by a \textit{local} uniaxial anisotropy with an estimated energy barrier of the order of 4 meV. 
This scenario is consistent with the phenomenological description of CoPhOMe as a non-collinear spin-spiral, 
with local anisotropy easy axes tilted with respect to the helix axis. 

We have proposed a spin Hamiltonian, Eq~\eqref{Ham_TD}, that enables the comparison of magnetic properties determined 
by \emph{ab-initio} calculations with experiments in a very efficient way, still keeping the rich complexity of CoPhOMe
(the same Hamiltonian without the anisotropy term describes also the thermodynamics of MnPhOMe).     
We have focused on the dependence of a characteristic energy scale $\Delta_\xi$ on the structural properties of CoPhOMe, parameterized by the angle $\theta$ formed by the local anisotropy axes of Co ions with the \emph{c} axis. We derived an analytic formula that describes the non-trivial dependence of $\Delta_\xi$ on the MAE, on the Co-NIT exchange coupling and on $\theta$. 
This proves the increased complexity of our thermodynamic model with respect to the Ising model -- commonly employed to rationalize SCMs -- 
in which it is trivially $\Delta_\xi=2J_{\rm ex}$. 

The relative strength of the Co-NIT exchange coupling with respect to the MAE is consistent with the formation of broad DWs as elementary excitations. 
In this sense, DFT results provide a new framework to describe the static and dynamic properties of CoPhOMe and solve the apparent contradiction of $\Delta_\xi$ being significantly different from $\Delta_\tau$ (barrier of the relaxation time): the latter is expected to be of the order of the DW energy $\mathcal{E}_{\rm dw}(\theta)$ and the former roughly $70\%$ of it. 

From a methodological perspective, the present study shows how the molecular approach spontaneously leads to isolate,
from an extended system, individual building blocks that can be studied with deeper accuracy, both theoretically and experimentally. 
The crucial information extracted from those building blocks and from their arrangement in the extended system is retained at a coarse-grained level, 
in a spin Hamiltonian through which \textit{equilibrium} and \textit{out-of-equilibrium} thermodynamics can -- in principle -- be modeled. 

From a more applicative point of view, CoPhOMe could serve as a playground to better understand how thermal fluctuations affect \textit{broad} DWs, typically hosted in ferromagnetic nanowires and described through a Hamiltonian\cite{Ar_Abanov_PRL_10a,Ar_Abanov_PRL_10b,B_Braun_AdvPhys_2012} equivalent to Eq.~\eqref{Ham_continuum_2}. Domain walls in nanowires have been proposed as magnetic-memory elements to be manipulated by electric current\cite{Parkin_Science_08,Hayashi_Science_08,Allenspach_PRL08,Tatara_PRL04}  
or injected spin waves\cite{Yan_11,Yan_12}. Moreover, a detailed characterization of the electronic properties of these magnetic molecular helices is also relevant to further understand the interplay between structural chirality and magnetism in these materials. For instance , this interplay may give rise to 
inverse magneto-chiral effect\cite{Wagniere_PRA_89},  i.e. magnetization induced by non-polarized light, or chiral-induced spin selectivity effects\cite{Goehler_Science11,Manso_NatComm14}, both of relevance for spintronics applications.


\begin{acknowledgments}
Work supported by Italian MIUR through FIRB project \emph{Nanomagneti molecolari su superfici metalliche e magnetiche per applicazioni nella spintronica molecolare} (RBAP117RWN). 
We acknowledge the CINECA Award No. HP10B2TGNW (2014), for the availability of high performance computing resources and support.
A. V.  would like to acknowledge Lorenzo Sorace for fruitful discussions and suggestions. 
We acknowledge the financial support of ETH Zurich and the Swiss National Science Foundation. 
\end{acknowledgments}


\appendix

\section{Continuum limit\label{Continuum-limit}}
Let us start considering the continuum limit for the M-NIT exchange interaction. 
Hamiltonian~\eqref{equationb} is treated at the same level of approximation as Eq.~\eqref{Ham_TD},  
namely Co spin operators are replaced by classical vectors $|\mathbf{S}_{2r}|=1$ and NIT spins by Pauli operators $\hat{\boldsymbol{\sigma}}_{2r+1}$: 
\begin{equation}
\mathcal{H}_{\text{M-NIT}}= -\sum_{r} \left[ 
J_\alpha\,\mathbf{S}_{2r}\cdot\hat{\boldsymbol{\sigma}}_{2r+1} + J_\beta\,\mathbf{S}_{2r}\cdot\hat{\boldsymbol{\sigma}}_{2r-1} \right]  
\label{Ham_AppA}	\,.
\end{equation}
For a given configuration of the two classical spins (representing metal ions) $\mathbf{S}_{2r}$ and $\mathbf{S}_{2r+2}$, the interlaying 
NIT spin experiences an exchange field $\mathbf{h}^{\rm ex}_{2r+1}= J_\alpha\mathbf{S}_{2r}+J_\beta\mathbf{S}_{2r+2}$. 
The corresponding eigenvalues $\pm|\mathbf{h}^{\rm ex}_{2r+1}|$ depend parametrically on the orientation of both $\mathbf{S}_{2r}$ and $\mathbf{S}_{2r+2}$. 
Let us first compute the squared modulus of the local exchange field: 
\begin{equation}
\begin{split}
|\mathbf{h}^{\rm ex}_{2r+1}|^2&= \left(J_\alpha\mathbf{S}_{2r}+J_\beta\mathbf{S}_{2r+2}\right)^2 \\
&=J_\alpha^2+J_\beta^2+2 J_\alpha J_\beta \mathbf{S}_{2r}\cdot\mathbf{S}_{2r+2} \\
&=J_\alpha^2+J_\beta^2+2 J_\alpha J_\beta - J_\alpha J_\beta\left( \mathbf{S}_{2r}-\mathbf{S}_{2r+2}\right)^2\\
&\simeq J_\alpha^2+J_\beta^2+2 J_\alpha J_\beta - J_\alpha J_\beta\left(\partial_z \mathbf{S}\right)^2
\end{split}
\end{equation}
where we used the fact that classical spins have unitary modulus and substituted the difference between successive classical spins by the spatial derivative.  
This approximation is essential to the continuum limit and relies on the assumption that large misalignment between $\mathbf{S}_{2r}$ and $\mathbf{S}_{2r+2}$ is highly unlikely,  because energetically not favorable. Consistently, a Taylor expansion with respect to this derivative can also be performed to obtain 
\begin{equation}
\begin{split}
|\mathbf{h}^{\rm ex}_{2r+1}| &= |J_\alpha+J_\beta|\sqrt{1-\rho \left(\partial_z \mathbf{S}\right)^2} \\
&\simeq |J_\alpha+J_\beta|\left[1-\frac{1}{2}\rho\left(\partial_z \mathbf{S}\right)^2\right]
\end{split}
\end{equation}
with $\rho=J_\alpha J_\beta/(J_\alpha+J_\beta)^2$, always smaller than one. 
Small deviations from the ground state of Hamiltonian~\eqref{Ham_AppA} are described by the following functional of classical spins 
\begin{equation}
\label{Ham_continuum_AppA}
\mathcal{H}_{\rm ex}=  \frac{|J_\alpha+J_\beta|}{2}\rho \int \left(\partial_z \mathbf{S}\right)^2 \, dz \,+\, \text{const.} 
\end{equation} 
\begin{figure}[t!]
\begin{center}
\includegraphics[width=1\linewidth]{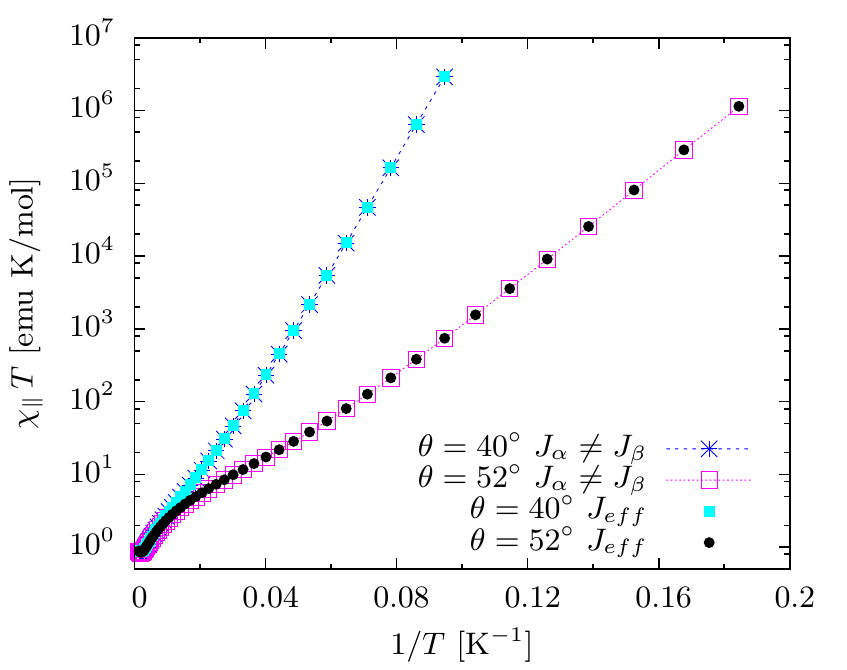}  
\end{center}
\caption{Color online. $\chi_\parallel=\partial M_\parallel/\partial B_\parallel$ along the chain axis computed in $B=0$ for $\theta=40^\circ,\,52^\circ$. 
Line-plus-symbol curves are obtained using the DFT values (set $a$) for $J_\alpha\ne J_\beta$; symbols correspond to calculations performed with 
a single, \textit{effective} value of $J_{\rm ex}= 2 J_\alpha J_\beta /(J_\alpha + J_\beta)$ (see the text).  \label{two_J}} 
\end{figure} 
Note that for $J_\alpha=J_\beta=J_{\rm ex}$ it is $\rho=1/4$ and the exchange contribution given in Hamiltonian \eqref{Ham_continuum_2} is obtained.  
Moreover, as long as the continuum approximation is legitimate, one expects to obtain the same results using either $J_\alpha\ne J_\beta$ or 
an effective unique coupling $J_{\rm eff}= 2 J_\alpha J_\beta /(J_\alpha + J_\beta)$. We have seen that the set $a$ of DFT calculations yields $J_\alpha=-404$ K and $J_\beta=-622$ K and that both  are  significantly larger than the estimate obtained for $D_z$. 
Therefore, the continuum limit should be justified for the set of parameters considered in this work\cite{Billoni2011}.   
If -- in turn -- \textit {all} the excitations that are relevant for thermodynamics are contained in Hamiltonian \eqref{Ham_continuum_2}, 
it should be equivalent to compute, e.g., the susceptibility  using two different coupling  $J_\alpha=-404$ K and $J_\beta=-622$ K or a unique effective coupling. 
This expectation is indeed confirmed quantitatively by Fig.~\ref{two_J} where curves computed using the two different values for 
$J_\alpha$ and $J_\beta$ given above or $J_{\rm eff}= 2 J_\alpha J_\beta /(J_\alpha + J_\beta)\simeq-490$ K are shown.
This justifies why a single value of the exchange coupling,  rounded to $J_{\rm ex}=-500$ K, was used to study thermodynamic properties.

In the following we show that in the continuum limit the anisotropy energy in Hamiltonian~\eqref{Ham_TD} transforms into the \textit{uniaxial} anisotropy term of the functional~\eqref{Ham_continuum_2}. Let us start rewriting the anisotropy energy as 
\begin{equation}
\label{Ham_ani_AppA}
\begin{split}
\mathcal{H}_{\rm ani}&= - D_z\int \sum_{p,r} \left(\mathbf{e}_r\cdot\mathbf{S}\right)^2 \delta(z-p-r) dz \\
&=-D_z \int \sum_{p,r} \left\{\sin^2\theta\left[\cos^2\varphi(S^x)^2+\sin^2\varphi(S^y)^2\right] \right. \\
& \left. + \cos^2\theta(S^z)^2 \right\} \delta(z-p-r) dz
\end{split}
\end{equation} 
where $\mathbf{e}_r=(\sin\theta\cos\left(\varphi(r)\right),\sin\theta\sin\left(\varphi(r)\right),\cos\theta)$ is the director of each local anisotropy axis 
(remember that $\varphi=2\pi\,r/3$ imposed by the crystal symmetry) and $\delta(\dots)$ is the Dirac delta function, assuming unitary lattice units. 
As $|J_{\rm ex}|$ is much larger than $D_z$, it is reasonable to assume that the spin profile varies smoothly in low-energy configurations. Then, similarly to what is usually done 
for spin spirals~\cite{Bode2007}, the $\cos^2\varphi$ and $\sin^2\varphi$ in Eq.~\eqref{Ham_ani_AppA} 
can be replaced by their spatial averages, namely $\langle\cos^2\varphi\rangle$=$\langle\sin^2\varphi\rangle$=$1/2$ 
(this is equivalent to assuming \textit{a priori} that DW excitations extend over several lattice units). 
The dependence on the site index $(p+r)$ thus drops from Eq.~\eqref{Ham_ani_AppA} as well as the summation and the Dirac delta function. 
More explicitly:  
\begin{equation}
\label{Ham_ani_AppA_1}
\begin{split}
\mathcal{H}_{\rm ani}=&-D_z \int \left\{\frac{1}{2}\sin^2\theta\left[(S^x)^2+(S^y)^2\right] \right. \\
& \left. + \cos^2\theta(S^z)^2 \right\}  dz \\
=&-D_z \int \left\{\frac{1}{2}\sin^2\theta\left[1-(S^z)^2\right] \right. \\
& \left. + \cos^2\theta(S^z)^2 \right\}  dz \\
=&-D_z \int \frac{1}{2}\sin^2\theta \, dz \\
&-D_z \int \left[\cos^2\theta -\frac{1}{2}\sin^2\theta\right] (S^z)^2 \, dz 
\end{split}
\end{equation}
where we have used the fact that $(S^x)^2+(S^y)^2=1-(S^z)^2$. 
The second last line in Eq.~\eqref{Ham_ani_AppA_1} gives a contribution that does not depend on the orientation of 
the classical spins $\mathbf{S}(z)$; however, this term will generally depend on the crystal structure, through $\theta$. 
The last line gives the anisotropy contribution reported in Hamiltonian~\eqref{Ham_continuum_2}.   
The fact that the effective anisotropy  
$D_z \left(3\cos^2\theta-1\right) /2$ is  smaller than $D_z$ for any $\theta\ne0$ and the agreement between 
the analytic model and transfer-matrix results reported in Fig.~\ref{Fig_Delta_xi} justify \textit{a posteriori} the validity of the approximations made to 
obtain  Hamiltonian \eqref{Ham_continuum_2} and the expression for the DW energy of CoPhOMe: $\mathcal{E}_{\rm dw}(\theta)=\sqrt{2 |J_{\rm ex}|D_z \left(3 \cos^2\theta -1\right)}$.

\section{Transfer Matrix method\label{Transfer-Matrix}} 
The thermodynamic properties of classical-spin chains with nearest-neighbor interactions can be computed very efficiently with the (numerical) transfer-matrix technique~\cite{Blume75PRB,Tannous}. This method can be extended to models in which classical and quantum spins alternate, like in our case\cite{gatteschi2013}.    
Noting that the quantum-spin operators are not directly coupled with each other, one can integrate out their degrees of freedom independently. 
Referring to Hamiltonian~\eqref{Ham_TD}, we add the Zeeman contribution due to $\mathbf{B}$ to the exchange field  acting on a generic quantum spin 
(introduced in Appendix~\ref{Continuum-limit}), which yields a total field $\mathbf{h}_{p,2r+1}=\mathbf{h}^{\rm ex}_{p,2r+1} + \mu_B g \mathbf{B}$. 
This field depends \textit{parametrically} on the orientation of the two classical spins $\mathbf{S}_{p,2r}$ and $\mathbf{S}_{p,2r+2}$.  Note that the 
cell index $p$ has been reintroduced, in accordance with Hamiltonian~\eqref{Ham_TD}. 
The trace over the degrees of freedom of the quantum spin $\boldsymbol{\sigma}_{p,2r+1}$ just brings a contribution $2\cosh\left(\beta|\mathbf{h}_{p,2r+1}|\right)$ 
into the partition function  (here it is $\beta=1/T$ not to be confused with the label for the exchange coupling used in the rest of the paper).  
The total partition function reads
\begin{equation}
\label{TM_kernel_trace}
\begin{split}
\mathcal{Z}&=\int \Pi_{p,r} d\Omega_{p,2r} 
\mathcal{K} (\mathbf{S}_{1,2},\, \mathbf{S}_{1,4})\,
\mathcal{K} (\mathbf{S}_{1,4},\, \mathbf{S}_{1,6})\, \\
&\times\mathcal{K} (\mathbf{S}_{1,6},\, \mathbf{S}_{2,2})\, 
\mathcal{K} (\mathbf{S}_{2,2},\, \mathbf{S}_{2,4})\, \ldots \,
\mathcal{K}(\mathbf{S}_{N_p,6},\, \mathbf{S}_{1,2})\,,
\end{split}
\end{equation}
where \textit{elementary} kernels\cite{Wyld} are defined as  
\begin{equation}
\label{M_kernel}
\begin{split}
\mathcal{K}&(\mathbf{S}_{p,2r},\, \mathbf{S}_{p,2r+2})=2\cosh\left(\beta|\mathbf{h}_{p,2r+1}|\right)\\	
&\times\exp\left(\beta\mathbf{S}_{p,2r}\widetilde{D}_{2r}\mathbf{S}_{p,2r}\right)				
\times\exp\left(\beta\mu_B G \mathbf{B}\cdot\mathbf{S}_{p,2r}\right)		\,.				
\end{split}
\end{equation}
Generally, the kernels $\mathcal{K} (\mathbf{S}_{p,2},\, \mathbf{S}_{p,4})$, $\mathcal{K} (\mathbf{S}_{p,4},\, \mathbf{S}_{p,6})$,  
$\mathcal{K} (\mathbf{S}_{1,6},\, \mathbf{S}_{2,2})$ are not equivalent, which mirrors the fact that the three Co atoms in each cell 
are not equivalent from the magnetic point of view. This is crucial to model reciprocal non-collinearity of anisotropy axes. 
A kernel that involves only equivalent spins can, however, be obtained by tracing over the degrees of freedom of two Co spins in a generic cell, for instance those labeled with 4 and 6:  
\begin{equation}
\label{Three-fold_kernel}
\begin{split}
\mathcal{K}_{\rm TM}&(\mathbf{S}_{p,2},\, \mathbf{S}_{p+1,2})=
\int  d\Omega_{p,4}\int\mathcal{K}(\mathbf{S}_{p,2},\,\mathbf{S}_{p,4}) \\
&\times\mathcal{K}(\mathbf{S}_{p,4},\,\mathbf{S}_{p,6})\,\mathcal{K}(\mathbf{S}_{p,6},\,\mathbf{S}_{p+1,2})\,d\Omega_{p,6} \,.
\end{split} 
\end{equation}
Then, the calculation of the partition function in Eq.~\eqref{TM_kernel_trace} reduces to the following eigenvalue problem 
\begin{equation}
\label{TM_integral_eq}
\int
\mathcal{K}_{\rm TM} (\mathbf{S}_{i},\,\mathbf{S}_{i+1})W_{m}(\mathbf{S}_{i+1})
d\Omega_{i+1} =\lambda_{m} W_{m}(\mathbf{S}_{i}) \,,
\end{equation}
whose eigenvalues may typically be ordered from the largest to the smallest one: $\lambda_{0}>\lambda_{1}>\lambda_{2}>\ldots $. 
The integrals on the solid angles were performed numerically  by discretizing the unitary sphere~\cite{Stroud,McLaren,Abramowitz}.  
As a result, the eigenvalue problem~\eqref{TM_integral_eq} was converted into a linear-algebra problem\cite{Vindigni06APA}. 
The resulting matrix -- to be diagonalized -- is generally not symmetric and its eigenvalues are pairs of complex conjugates. 
In the thermodynamic limit $N_p\rightarrow\infty$ the partition function equals the  $N_p$-th power of the largest eigenvalue,   
$\mathcal{Z}=\lambda_0^{N_p}$ ($N_p$ is the number of cells); therefore $\lambda_0$ has to be real. For each temperature, the number of points 
used to sample solid angles was increased until the free-energy $F= -T \ln\left(\mathcal{Z}\right)$ 
converged to a stable value\cite{Sangiorgio2014}. From the free-energy, the magnetization per unit cell $M_\eta=-\partial F / (N_p \partial B_\eta)$  
and the susceptibility $\chi_\eta= \partial M_\eta /\partial B_\eta$ were obtained ($\eta=x,y,z$ indicates the crystallographic axis along which $\mathbf{B}$ is applied).

\section{\label{DW_diffusion}Glauber model and diffusion of broad domain walls}  
One of the reasons for the fortune of the Glauber model in the SCM community certainly  lies on the possibility of deriving analytic results. 
On the contrary, the treatment  of stochastic dynamics in the framework of the Heisenberg model is not as simple and even numerical simulations -- involving a physically meaningful time scale -- 
are challenging. The starting point is the stochastic  Landau-Lifshitz-Gilbert equation, which can be integrated following the Langevin  or the equivalent time-quantified Monte-Carlo approach\cite{Billoni2011,Hinzke00PRB}.   
Here we limit ourselves to propose a mapping of the elementary time scale of the Glauber model into the diffusion coefficient associated with the motion of a broad DW.  
Within the Glauber model the probability to flip a spin occupying the site $i$ of an Ising chain is given by\cite{glauber1963} 
\begin{equation}
\label{Glauber_w}
w_i =\frac{\alpha}{2} \left[1-\frac{\gamma}{2}\sigma_i\left(\sigma_{i+1}+\sigma_{i-1}\right) \right]\,,
\end{equation}
where $\sigma_i=\pm 1$ are Ising spin variables and $\gamma=\tanh\left(2J_{\rm ex}/T\right)$. 
When the $i$-th spin is the front edge of a DW, it is $\sigma_{i+1}+\sigma_{i-1}=0$; therefore, $\alpha/2$ can be identified with the probability to move a DW of one site per unit time. 
Let us define $P(z,t)$ as the probability to find a DW at the coordinate $z$ -- being now a continuous variable --  at time $t$. 
This probability obeys the master equation   
\begin{equation}
\label{P_equation_1}
\partial_t P(z,t)= \frac{\alpha}{2} \left[P(z+\delta,t) +P(z-\delta,t) -2P(z,t)  \right] 
\end{equation}
where $\delta$ is the elementary displacement of the stochastic process. By expanding 
Eq.~\eqref{P_equation_1} for small $\delta$,  the standard 1D diffusion equation is recovered: 
\begin{equation}
\label{P_equation_2}
\partial_t P(z,t)=\frac{\alpha}{2} \delta^2 \partial^2_z P(z,t)  
\end{equation}
in which $D=\alpha \delta^2/2$ plays the role of diffusion coefficient. Assuming that a DW covers a distance of the order of the correlation length within a lapse of time equal to the relaxation time,  
one gets to the equivalence $\xi^2\simeq D \tau$. As already reported in Ref.~\onlinecite{Coulon06Springer}, for the Ising limit corresponding to $\delta=1$, this mapping yields the same asymptotic behavior  
obtained by Glauber: $\tau_\infty=2\xi^2/\alpha$. The subscript indicates that this result actually holds for the infinite chain. For a finite chain, still in the framework of Glauber dynamics, it can be shown that the low-temperature behavior of  the relaxation time is $\tau\simeq L \,{\rm e}^{\mathcal{E}_{\rm dw}/T}/(2\alpha)$, $L$ being the chain length\cite{daSilva_PRE_95}. 
Our goal is now to estimate what effective $\alpha$ should be used to mimic the diffusion of a \textit{broad} DW with Glauber dynamics.  
The first step is to identify $\delta$ with the DW width $w$; for the Hamiltonian~\eqref{Ham_continuum_2}, the latter reads $w= \sqrt{|J_{\rm ex}|/\left[2D_z (3\cos^2\theta-1)\right]}$.  
Adapting the numerical factors given in Ref.~\onlinecite{Billoni2011,gatteschi2013} to Eq.~\eqref{P_equation_2}, the diffusion coefficient can be related to the
spin-Hamiltonian parameters: 
\begin{equation}
\label{Diff_coeff}
D\simeq 0.34 \,\frac{w^2}{\tau_{\rm d}} \frac{T}{\mathcal{E}_{\rm dw}(\theta)} 
\end{equation}
where $\tau_{\rm d}$ is the damping time $\tau_{\rm d} \simeq \hbar/(2D_z\alpha_\text{G})$, $\alpha_\text{G}$ is the Gilbert damping\cite{Gilbert_2004} and 
$D_z$ is the anisotropy on each Co.~\cite{Billoni2011,gatteschi2013}. 
For this special case the attempt frequency of the Glauber model and the diffusion coefficient are related by the equation $\alpha =2D /w^2$. The low-temperature expansion given above for the relaxation time 
of a finite Ising chain can \textit{tentatively} be extended to a chain with broad DWs: 
\begin{equation}
\label{tau_eff}
\tau \simeq \,\frac{L}{w} \frac{ {\rm e}^\kappa} {2\alpha} = \frac{L}{2} \frac{\tau_{\rm d}}{w}  \frac{  \kappa\,{\rm e}^\kappa   }{0.68}
\simeq \frac{L}{2} \left(\frac{\tau_{\rm d}}{w}  \frac{6}{0.68} \right){\rm e}^{1.06 \kappa }\,;
\end{equation}
in the last passage the approximation $\kappa\,{\rm e}^\kappa\simeq 6\,{\rm e}^{1.06 \kappa } $ was made based on the fact that the experimentally relevant range of $\kappa=\mathcal{E}_{\rm dw}(\theta)/T$ 
is $12 < \kappa < 22$. In Eq.~\eqref{tau_eff}, the inverse of the term in parenthesis represents the effective Glauber attempt frequency that should emulate the diffusion of a broad DW. 
For our computation parameters and $50^\circ \leq \theta \leq 54^\circ$, it is $\alpha_{\rm eff} \sim 10^{15}\alpha_\text{G}$ s$^{-1}$.  
For the temperature range where slow relaxation is observed in CoPhOMe, the order of magnitude of the Gilbert damping can be roughly estimated from EPR spectra of the CoNIT$_2$, shown in Ref.~\onlinecite{caneschi2002}. 
In a trivial precessional model, the half-with-half-maximum $\Delta H$ of the resonance peak would be related to the  Gilbert damping as $\gamma_{\rm g} \Delta H\simeq \alpha_\text{G}\,\omega_{\rm L}$,
with $\omega_{\rm L}=\gamma_{\rm g} H_0$ being the Larmor frequency, $H_0$ the resonance field and $\gamma_{\rm g}$ the gyromagnetic ratio. Then one obtains  
$\alpha_\text{G}\simeq\Delta H/ H_0\simeq 0.15$, from which $\alpha_{\rm eff} \sim 1.5 \cdot 10^{14}$ s$^{-1}$. 
To simulate the data reported in Ref.~\onlinecite{BoganiPRL2004} the Glauber model 
with $\alpha=2.6\cdot 10^{13}$ s$^{-1}$ was used.  
Considering the crude approximation made to estimate $\alpha_\text{G}$ and that the parameters in Hamiltonian~\eqref{Ham_TD} have been deduced from DFT calculations and not fitted, the present mapping sounds in reasonable agreement with previously published results.   
  

\begin{thebibliography}{70}%
\makeatletter
\providecommand \@ifxundefined [1]{%
 \@ifx{#1\undefined}
}%
\providecommand \@ifnum [1]{%
 \ifnum #1\expandafter \@firstoftwo
 \else \expandafter \@secondoftwo
 \fi
}%
\providecommand \@ifx [1]{%
 \ifx #1\expandafter \@firstoftwo
 \else \expandafter \@secondoftwo
 \fi
}%
\providecommand \natexlab [1]{#1}%
\providecommand \enquote  [1]{``#1''}%
\providecommand \bibnamefont  [1]{#1}%
\providecommand \bibfnamefont [1]{#1}%
\providecommand \citenamefont [1]{#1}%
\providecommand \href@noop [0]{\@secondoftwo}%
\providecommand \href [0]{\begingroup \@sanitize@url \@href}%
\providecommand \@href[1]{\@@startlink{#1}\@@href}%
\providecommand \@@href[1]{\endgroup#1\@@endlink}%
\providecommand \@sanitize@url [0]{\catcode `\\12\catcode `\$12\catcode
  `\&12\catcode `\#12\catcode `\^12\catcode `\_12\catcode `\%12\relax}%
\providecommand \@@startlink[1]{}%
\providecommand \@@endlink[0]{}%
\providecommand \url  [0]{\begingroup\@sanitize@url \@url }%
\providecommand \@url [1]{\endgroup\@href {#1}{\urlprefix }}%
\providecommand \urlprefix  [0]{URL }%
\providecommand \Eprint [0]{\href }%
\@ifxundefined \urlstyle {%
  \providecommand \doi  [0]{\begingroup \@sanitize@url \@doi}%
  \providecommand \@doi [1]{\endgroup \@@startlink {\doibase
  #1}doi:\discretionary {}{}{}#1\@@endlink }%
}{%
  \providecommand \doi  [0]{doi:\discretionary{}{}{}\begingroup
  \urlstyle{rm}\Url }%
}%
\providecommand \doibase [0]{http://dx.doi.org/}%
\providecommand \Doi [0]{\begingroup \@sanitize@url \@Doi }%
\providecommand \@Doi  [1]{\endgroup\@@startlink{\doibase#1}\@@Doi}%
\providecommand \@@Doi [1]{#1\@@endlink}%
\providecommand \selectlanguage [0]{\@gobble}%
\providecommand \bibinfo  [0]{\@secondoftwo}%
\providecommand \bibfield  [0]{\@secondoftwo}%
\providecommand \translation [1]{[#1]}%
\providecommand \BibitemOpen [0]{}%
\providecommand \bibitemStop [0]{}%
\providecommand \bibitemNoStop [0]{.\EOS\space}%
\providecommand \EOS [0]{\spacefactor3000\relax}%
\providecommand \BibitemShut  [1]{\csname bibitem#1\endcsname}%
\bibitem [{\citenamefont {Gatteschi}\ \emph {et~al.}(2006)\citenamefont
  {Gatteschi}, \citenamefont {Sessoli},\ and\ \citenamefont
  {Villain}}]{gatteschi2006}%
  \BibitemOpen
  \bibfield  {author} {\bibinfo {author} {\bibfnamefont {D.}~\bibnamefont
  {Gatteschi}}, \bibinfo {author} {\bibfnamefont {R.}~\bibnamefont {Sessoli}},
  \ and\ \bibinfo {author} {\bibfnamefont {J.}~\bibnamefont {Villain}},\ }\href
  {http://books.google.it/books?id=PvZ9drxKW0oC} {\emph {\bibinfo {title}
  {Molecular Nanomagnets}}},\ Mesoscopic Physics and Nanotechnology\ (\bibinfo
  {publisher} {OUP Oxford},\ \bibinfo {year} {2006})\ ISBN \bibinfo {isbn}
  {9780198567530}\BibitemShut {NoStop}%
\bibitem [{\citenamefont {Bartolom\'e}\ \emph {et~al.}(2014)\citenamefont
  {Bartolom\'e}, \citenamefont {Fernando},\ and\ \citenamefont
  {Fern\'andez}}]{Bartolome2014}%
  \BibitemOpen
  \bibfield  {author} {\bibinfo {author} {\bibfnamefont {J.}~\bibnamefont
  {Bartolom\'e}}, \bibinfo {author} {\bibfnamefont {L.}~\bibnamefont
  {Fernando}}, \ and\ \bibinfo {author} {\bibfnamefont {J.~F.}\ \bibnamefont
  {Fern\'andez}},\ }\Doi {10.1007/978-3-642-40609-6} {\emph {\bibinfo {title}
  {Molecular Magnets}}},\ Physics and Applications\ (\bibinfo  {publisher}
  {Springer-Verlag},\ \bibinfo {address} {Berlin Heidelberg},\ \bibinfo {year}
  {2014})\ ISBN \bibinfo {isbn} {978-3-642-40609-6}\BibitemShut {NoStop}%
\bibitem [{\citenamefont {Sessoli}\ \emph {et~al.}(1993)\citenamefont
  {Sessoli}, \citenamefont {Gatteschi}, \citenamefont {Caneschi},\ and\
  \citenamefont {Novak}}]{sessoli1993}%
  \BibitemOpen
  \bibfield  {author} {\bibinfo {author} {\bibfnamefont {R.}~\bibnamefont
  {Sessoli}}, \bibinfo {author} {\bibfnamefont {D.}~\bibnamefont {Gatteschi}},
  \bibinfo {author} {\bibfnamefont {A.}~\bibnamefont {Caneschi}}, \ and\
  \bibinfo {author} {\bibfnamefont {M.~A.}\ \bibnamefont {Novak}},\ }\href
  {http://dx.doi.org/10.1038/365141a0} {\bibfield  {journal} {\bibinfo
  {journal} {Nature},\ }\textbf {\bibinfo {volume} {365}},\ \bibinfo {pages}
  {141} (\bibinfo {year} {1993})}\BibitemShut {NoStop}%
\bibitem [{\citenamefont {Wernsdorfer}\ and\ \citenamefont
  {Sessoli}(1999)}]{Wernsdorfer_Science99}%
  \BibitemOpen
  \bibfield  {author} {\bibinfo {author} {\bibfnamefont {W.}~\bibnamefont
  {Wernsdorfer}}\ and\ \bibinfo {author} {\bibfnamefont {R.}~\bibnamefont
  {Sessoli}},\ }\Doi {10.1126/science.284.5411.133} {\bibfield  {journal}
  {\bibinfo  {journal} {Science},\ }\textbf {\bibinfo {volume} {284}},\
  \bibinfo {pages} {133} (\bibinfo {year} {1999})}\BibitemShut {NoStop}%
\bibitem [{\citenamefont {Luzon}\ \emph {et~al.}(2008)\citenamefont {Luzon},
  \citenamefont {Bernot}, \citenamefont {Hewitt}, \citenamefont {Anson},
  \citenamefont {Powell},\ and\ \citenamefont {Sessoli}}]{LuzonPRL2008}%
  \BibitemOpen
  \bibfield  {author} {\bibinfo {author} {\bibfnamefont {J.}~\bibnamefont
  {Luzon}}, \bibinfo {author} {\bibfnamefont {K.}~\bibnamefont {Bernot}},
  \bibinfo {author} {\bibfnamefont {I.~J.}\ \bibnamefont {Hewitt}}, \bibinfo
  {author} {\bibfnamefont {C.~E.}\ \bibnamefont {Anson}}, \bibinfo {author}
  {\bibfnamefont {A.~K.}\ \bibnamefont {Powell}}, \ and\ \bibinfo {author}
  {\bibfnamefont {R.}~\bibnamefont {Sessoli}},\ }\Doi
  {10.1103/PhysRevLett.100.247205} {\bibfield  {journal} {\bibinfo  {journal}
  {Phys. Rev. Lett.},\ }\textbf {\bibinfo {volume} {100}},\ \bibinfo {pages}
  {247205} (\bibinfo {year} {2008})}\BibitemShut {NoStop}%
\bibitem [{\citenamefont {Sorace}\ \emph {et~al.}(2003)\citenamefont {Sorace},
  \citenamefont {Wernsdorfer}, \citenamefont {Thirion}, \citenamefont {Barra},
  \citenamefont {Pacchioni}, \citenamefont {Mailly},\ and\ \citenamefont
  {Barbara}}]{Sorace_PRB_03}%
  \BibitemOpen
  \bibfield  {author} {\bibinfo {author} {\bibfnamefont {L.}~\bibnamefont
  {Sorace}}, \bibinfo {author} {\bibfnamefont {W.}~\bibnamefont {Wernsdorfer}},
  \bibinfo {author} {\bibfnamefont {C.}~\bibnamefont {Thirion}}, \bibinfo
  {author} {\bibfnamefont {A.-L.}\ \bibnamefont {Barra}}, \bibinfo {author}
  {\bibfnamefont {M.}~\bibnamefont {Pacchioni}}, \bibinfo {author}
  {\bibfnamefont {D.}~\bibnamefont {Mailly}}, \ and\ \bibinfo {author}
  {\bibfnamefont {B.}~\bibnamefont {Barbara}},\ }\Doi
  {10.1103/PhysRevB.68.220407} {\bibfield  {journal} {\bibinfo  {journal}
  {Phys. Rev. B},\ }\textbf {\bibinfo {volume} {68}},\ \bibinfo {pages}
  {220407} (\bibinfo {year} {2003})}\BibitemShut {NoStop}%
\bibitem [{\citenamefont {Bogani}\ and\ \citenamefont
  {Wernsdorfer}(2008)}]{Bogani_NatMat08}%
  \BibitemOpen
  \bibfield  {author} {\bibinfo {author} {\bibfnamefont {L.}~\bibnamefont
  {Bogani}}\ and\ \bibinfo {author} {\bibfnamefont {W.}~\bibnamefont
  {Wernsdorfer}},\ }\Doi {10.1038/nmat2133} {\bibfield  {journal} {\bibinfo
  {journal} {Nature Mater.},\ }\textbf {\bibinfo {volume} {7}},\ \bibinfo
  {pages} {179} (\bibinfo {year} {2008})}\BibitemShut {NoStop}%
\bibitem [{\citenamefont {Bogani}\ \emph {et~al.}(2008)\citenamefont {Bogani},
  \citenamefont {Vindigni}, \citenamefont {Sessoli},\ and\ \citenamefont
  {Gatteschi}}]{bogani2008}%
  \BibitemOpen
  \bibfield  {author} {\bibinfo {author} {\bibfnamefont {L.}~\bibnamefont
  {Bogani}}, \bibinfo {author} {\bibfnamefont {A.}~\bibnamefont {Vindigni}},
  \bibinfo {author} {\bibfnamefont {R.}~\bibnamefont {Sessoli}}, \ and\
  \bibinfo {author} {\bibfnamefont {D.}~\bibnamefont {Gatteschi}},\ }\Doi
  {10.1039/B807824F} {\bibfield  {journal} {\bibinfo  {journal} {J. Mater.
  Chem.},\ }\textbf {\bibinfo {volume} {18}},\ \bibinfo {pages} {4750}
  (\bibinfo {year} {2008})}\BibitemShut {NoStop}%
\bibitem [{\citenamefont {Gatteschi}\ and\ \citenamefont
  {Vindigni}(2014)}]{gatteschi2013}%
  \BibitemOpen
  \bibfield  {author} {\bibinfo {author} {\bibfnamefont {D.}~\bibnamefont
  {Gatteschi}}\ and\ \bibinfo {author} {\bibfnamefont {A.}~\bibnamefont
  {Vindigni}},\ }in\ \href@noop {} {\emph {\bibinfo {booktitle} {Molecular
  Magnets}}},\ \bibinfo {editor} {edited by\ \bibinfo {editor} {\bibfnamefont
  {J.}~\bibnamefont {Bartolom\'e}}, \bibinfo {editor} {\bibfnamefont
  {L.}~\bibnamefont {Fernando}}, \ and\ \bibinfo {editor} {\bibfnamefont
  {J.~F.}\ \bibnamefont {Fern\'andez}}}\ (\bibinfo  {publisher}
  {Springer-Verlag},\ \bibinfo {address} {Berlin Heidelberg},\ \bibinfo {year}
  {2014})\BibitemShut {NoStop}%
\bibitem [{\citenamefont {Steiner}\ \emph {et~al.}(1976)\citenamefont
  {Steiner}, \citenamefont {Villain},\ and\ \citenamefont
  {Windsor}}]{SVW_AdvPhys76}%
  \BibitemOpen
  \bibfield  {author} {\bibinfo {author} {\bibfnamefont {M.}~\bibnamefont
  {Steiner}}, \bibinfo {author} {\bibfnamefont {J.}~\bibnamefont {Villain}}, \
  and\ \bibinfo {author} {\bibfnamefont {C.}~\bibnamefont {Windsor}},\ }\Doi
  {10.1080/00018737600101372} {\bibfield  {journal} {\bibinfo  {journal}
  {Advances in Physics},\ }\textbf {\bibinfo {volume} {25}},\ \bibinfo {pages}
  {87} (\bibinfo {year} {1976})}\BibitemShut {NoStop}%
\bibitem [{\citenamefont {De~Jongh}\ and\ \citenamefont
  {Miedema}(2001)}]{DeJongh-Miedema_AdvPhys01}%
  \BibitemOpen
  \bibfield  {author} {\bibinfo {author} {\bibfnamefont {L.~J.}\ \bibnamefont
  {De~Jongh}}\ and\ \bibinfo {author} {\bibfnamefont {A.~R.}\ \bibnamefont
  {Miedema}},\ }\Doi {10.1080/00018730110101412} {\bibfield  {journal}
  {\bibinfo  {journal} {Advances in Physics},\ }\textbf {\bibinfo {volume}
  {50}},\ \bibinfo {pages} {947} (\bibinfo {year} {2001})}\BibitemShut
  {NoStop}%
\bibitem [{\citenamefont {Sachdev}(2000)}]{Sachdev_Science_00}%
  \BibitemOpen
  \bibfield  {author} {\bibinfo {author} {\bibfnamefont {S.}~\bibnamefont
  {Sachdev}},\ }\Doi {10.1126/science.288.5465.475} {\bibfield  {journal}
  {\bibinfo  {journal} {Science},\ }\textbf {\bibinfo {volume} {288}},\
  \bibinfo {pages} {475} (\bibinfo {year} {2000})}\BibitemShut {NoStop}%
\bibitem [{\citenamefont {Coldea}\ \emph {et~al.}(2010)\citenamefont {Coldea},
  \citenamefont {Tennant}, \citenamefont {Wheeler}, \citenamefont {Wawrzynska},
  \citenamefont {Prabhakaran}, \citenamefont {Telling}, \citenamefont
  {Habicht}, \citenamefont {Smeibidl},\ and\ \citenamefont
  {Kiefer}}]{Coldea_Science_10}%
  \BibitemOpen
  \bibfield  {author} {\bibinfo {author} {\bibfnamefont {R.}~\bibnamefont
  {Coldea}}, \bibinfo {author} {\bibfnamefont {D.~A.}\ \bibnamefont {Tennant}},
  \bibinfo {author} {\bibfnamefont {E.~M.}\ \bibnamefont {Wheeler}}, \bibinfo
  {author} {\bibfnamefont {E.}~\bibnamefont {Wawrzynska}}, \bibinfo {author}
  {\bibfnamefont {D.}~\bibnamefont {Prabhakaran}}, \bibinfo {author}
  {\bibfnamefont {M.}~\bibnamefont {Telling}}, \bibinfo {author} {\bibfnamefont
  {K.}~\bibnamefont {Habicht}}, \bibinfo {author} {\bibfnamefont
  {P.}~\bibnamefont {Smeibidl}}, \ and\ \bibinfo {author} {\bibfnamefont
  {K.}~\bibnamefont {Kiefer}},\ }\Doi {10.1126/science.1180085} {\bibfield
  {journal} {\bibinfo  {journal} {Science},\ }\textbf {\bibinfo {volume}
  {327}},\ \bibinfo {pages} {177} (\bibinfo {year} {2010})}\BibitemShut
  {NoStop}%
\bibitem [{\citenamefont {Simon}\ \emph {et~al.}(2011)\citenamefont {Simon},
  \citenamefont {Bakr}, \citenamefont {Ma}, \citenamefont {Tai}, \citenamefont
  {Preiss},\ and\ \citenamefont {Greiner}}]{Simon_Nature_11}%
  \BibitemOpen
  \bibfield  {author} {\bibinfo {author} {\bibfnamefont {J.}~\bibnamefont
  {Simon}}, \bibinfo {author} {\bibfnamefont {W.~S.}\ \bibnamefont {Bakr}},
  \bibinfo {author} {\bibfnamefont {R.}~\bibnamefont {Ma}}, \bibinfo {author}
  {\bibfnamefont {M.~E.}\ \bibnamefont {Tai}}, \bibinfo {author} {\bibfnamefont
  {P.~M.}\ \bibnamefont {Preiss}}, \ and\ \bibinfo {author} {\bibfnamefont
  {M.}~\bibnamefont {Greiner}},\ }\Doi {10.1038/nature09994} {\bibfield
  {journal} {\bibinfo  {journal} {Nature},\ }\textbf {\bibinfo {volume}
  {472}},\ \bibinfo {pages} {307} (\bibinfo {year} {2011})}\BibitemShut
  {NoStop}%
\bibitem [{\citenamefont {Cl{\'e}rac}\ \emph {et~al.}(2002)\citenamefont
  {Cl{\'e}rac}, \citenamefont {Miyasaka}, \citenamefont {Yamashita},\ and\
  \citenamefont {Coulon}}]{Clerac2002}%
  \BibitemOpen
  \bibfield  {author} {\bibinfo {author} {\bibfnamefont {R.}~\bibnamefont
  {Cl{\'e}rac}}, \bibinfo {author} {\bibfnamefont {H.}~\bibnamefont
  {Miyasaka}}, \bibinfo {author} {\bibfnamefont {M.}~\bibnamefont {Yamashita}},
  \ and\ \bibinfo {author} {\bibfnamefont {C.}~\bibnamefont {Coulon}},\ }\Doi
  {10.1021/ja0203115} {\bibfield  {journal} {\bibinfo  {journal} {Journal of
  the American Chemical Society},\ }\textbf {\bibinfo {volume} {124}},\
  \bibinfo {pages} {12837} (\bibinfo {year} {2002})}\BibitemShut {NoStop}%
\bibitem [{\citenamefont {Coulon}\ \emph {et~al.}(2006)\citenamefont {Coulon},
  \citenamefont {Miyasaka},\ and\ \citenamefont {Cl\'erac}}]{Coulon06Springer}%
  \BibitemOpen
  \bibfield  {author} {\bibinfo {author} {\bibfnamefont {C.}~\bibnamefont
  {Coulon}}, \bibinfo {author} {\bibfnamefont {H.}~\bibnamefont {Miyasaka}}, \
  and\ \bibinfo {author} {\bibfnamefont {R.}~\bibnamefont {Cl\'erac}},\
  }\href@noop {} {\bibfield  {journal} {\bibinfo  {journal} {Struct. Bond.},\
  }\textbf {\bibinfo {volume} {122}},\ \bibinfo {pages} {163} (\bibinfo {year}
  {2006})}\BibitemShut {NoStop}%
\bibitem [{\citenamefont {Bogani}\ \emph {et~al.}(2004)\citenamefont {Bogani},
  \citenamefont {A.~Caneschi}, \citenamefont {Fedi}, \citenamefont {Gatteschi},
  \citenamefont {Massi}, \citenamefont {Novak}, \citenamefont {Pini},
  \citenamefont {Rettori}, \citenamefont {Sessoli},\ and\ \citenamefont
  {Vindigni}}]{BoganiPRL2004}%
  \BibitemOpen
  \bibfield  {author} {\bibinfo {author} {\bibfnamefont {L.}~\bibnamefont
  {Bogani}}, \bibinfo {author} {\bibfnamefont {A.}~\bibnamefont {A.~Caneschi}},
  \bibinfo {author} {\bibfnamefont {M.}~\bibnamefont {Fedi}}, \bibinfo {author}
  {\bibfnamefont {D.}~\bibnamefont {Gatteschi}}, \bibinfo {author}
  {\bibfnamefont {M.}~\bibnamefont {Massi}}, \bibinfo {author} {\bibfnamefont
  {M.~A.}\ \bibnamefont {Novak}}, \bibinfo {author} {\bibfnamefont {M.~G.}\
  \bibnamefont {Pini}}, \bibinfo {author} {\bibfnamefont {A.}~\bibnamefont
  {Rettori}}, \bibinfo {author} {\bibfnamefont {S.}~\bibnamefont {Sessoli}}, \
  and\ \bibinfo {author} {\bibfnamefont {A.}~\bibnamefont {Vindigni}},\ }\Doi
  {10.1103/PhysRevLett.92.207204} {\bibfield  {journal} {\bibinfo  {journal}
  {Phys. Rev. Lett.},\ }\textbf {\bibinfo {volume} {92}},\ \bibinfo {pages}
  {207204} (\bibinfo {year} {2004})}\BibitemShut {NoStop}%
\bibitem [{\citenamefont {Coulon}\ \emph {et~al.}(2004)\citenamefont {Coulon},
  \citenamefont {Cl\'erac}, \citenamefont {Lecren}, \citenamefont
  {Wernsdorfer},\ and\ \citenamefont {Miyasaka}}]{CoulonPRB04}%
  \BibitemOpen
  \bibfield  {author} {\bibinfo {author} {\bibfnamefont {C.}~\bibnamefont
  {Coulon}}, \bibinfo {author} {\bibfnamefont {R.}~\bibnamefont {Cl\'erac}},
  \bibinfo {author} {\bibfnamefont {L.}~\bibnamefont {Lecren}}, \bibinfo
  {author} {\bibfnamefont {W.}~\bibnamefont {Wernsdorfer}}, \ and\ \bibinfo
  {author} {\bibfnamefont {H.}~\bibnamefont {Miyasaka}},\ }\Doi
  {10.1103/PhysRevB.69.132408} {\bibfield  {journal} {\bibinfo  {journal}
  {Phys. Rev. B},\ }\textbf {\bibinfo {volume} {69}},\ \bibinfo {pages}
  {132408} (\bibinfo {year} {2004})}\BibitemShut {NoStop}%
\bibitem [{\citenamefont {Vindigni}\ \emph {et~al.}(2005)\citenamefont
  {Vindigni}, \citenamefont {Rettori}, \citenamefont {Bogani}, \citenamefont
  {Caneschi}, \citenamefont {Gatteschi}, \citenamefont {Sessoli},\ and\
  \citenamefont {Novak}}]{VindigniAPL05}%
  \BibitemOpen
  \bibfield  {author} {\bibinfo {author} {\bibfnamefont {A.}~\bibnamefont
  {Vindigni}}, \bibinfo {author} {\bibfnamefont {A.}~\bibnamefont {Rettori}},
  \bibinfo {author} {\bibfnamefont {L.}~\bibnamefont {Bogani}}, \bibinfo
  {author} {\bibfnamefont {A.}~\bibnamefont {Caneschi}}, \bibinfo {author}
  {\bibfnamefont {D.}~\bibnamefont {Gatteschi}}, \bibinfo {author}
  {\bibfnamefont {R.}~\bibnamefont {Sessoli}}, \ and\ \bibinfo {author}
  {\bibfnamefont {M.~A.}\ \bibnamefont {Novak}},\ }\Doi
  {http://dx.doi.org/10.1063/1.2001160} {\bibfield  {journal} {\bibinfo
  {journal} {Applied Physics Letters},\ }\textbf {\bibinfo {volume} {87}},\
  \bibinfo {pages} {3} (\bibinfo {year} {2005})}\BibitemShut {NoStop}%
\bibitem [{\citenamefont {Caneschi}\ \emph {et~al.}(2001)\citenamefont
  {Caneschi}, \citenamefont {Gatteschi}, \citenamefont {Lalioti}, \citenamefont
  {Sangregorio}, \citenamefont {Sessoli}, \citenamefont {Venturi},
  \citenamefont {Vindigni}, \citenamefont {Rettori}, \citenamefont {Pini},\
  and\ \citenamefont {Novak}}]{caneschi2001}%
  \BibitemOpen
  \bibfield  {author} {\bibinfo {author} {\bibfnamefont {A.}~\bibnamefont
  {Caneschi}}, \bibinfo {author} {\bibfnamefont {D.}~\bibnamefont {Gatteschi}},
  \bibinfo {author} {\bibfnamefont {N.}~\bibnamefont {Lalioti}}, \bibinfo
  {author} {\bibfnamefont {C.}~\bibnamefont {Sangregorio}}, \bibinfo {author}
  {\bibfnamefont {R.}~\bibnamefont {Sessoli}}, \bibinfo {author} {\bibfnamefont
  {G.}~\bibnamefont {Venturi}}, \bibinfo {author} {\bibfnamefont
  {A.}~\bibnamefont {Vindigni}}, \bibinfo {author} {\bibfnamefont
  {A.}~\bibnamefont {Rettori}}, \bibinfo {author} {\bibfnamefont {M.~G.}\
  \bibnamefont {Pini}}, \ and\ \bibinfo {author} {\bibfnamefont {M.~A.}\
  \bibnamefont {Novak}},\ }\href@noop {} {\bibfield  {journal} {\bibinfo
  {journal} {Angewandte Chemie International Edition},\ }\textbf {\bibinfo
  {volume} {40}},\ \bibinfo {pages} {1760} (\bibinfo {year}
  {2001})}\BibitemShut {NoStop}%
\bibitem [{\citenamefont {Glauber}(1963)}]{glauber1963}%
  \BibitemOpen
  \bibfield  {author} {\bibinfo {author} {\bibfnamefont {R.~J.}\ \bibnamefont
  {Glauber}},\ }\href@noop {} {\bibfield  {journal} {\bibinfo  {journal}
  {Journal of mathematical physics},\ }\textbf {\bibinfo {volume} {4}},\
  \bibinfo {pages} {294} (\bibinfo {year} {1963})}\BibitemShut {NoStop}%
\bibitem [{\citenamefont {Coulon}\ \emph {et~al.}(2009)\citenamefont {Coulon},
  \citenamefont {Cl\'erac}, \citenamefont {Wernsdorfer}, \citenamefont
  {Colin},\ and\ \citenamefont {Miyasaka}}]{Coulon_PRL09}%
  \BibitemOpen
  \bibfield  {author} {\bibinfo {author} {\bibfnamefont {C.}~\bibnamefont
  {Coulon}}, \bibinfo {author} {\bibfnamefont {R.}~\bibnamefont {Cl\'erac}},
  \bibinfo {author} {\bibfnamefont {W.}~\bibnamefont {Wernsdorfer}}, \bibinfo
  {author} {\bibfnamefont {T.}~\bibnamefont {Colin}}, \ and\ \bibinfo {author}
  {\bibfnamefont {H.}~\bibnamefont {Miyasaka}},\ }\Doi
  {10.1103/PhysRevLett.102.167204} {\bibfield  {journal} {\bibinfo  {journal}
  {Phys. Rev. Lett.},\ }\textbf {\bibinfo {volume} {102}},\ \bibinfo {pages}
  {167204} (\bibinfo {year} {2009})}\BibitemShut {NoStop}%
\bibitem [{\citenamefont {Coulon}\ \emph {et~al.}(2007)\citenamefont {Coulon},
  \citenamefont {Cl\'erac}, \citenamefont {Wernsdorfer}, \citenamefont {Colin},
  \citenamefont {Saitoh}, \citenamefont {Motokawa},\ and\ \citenamefont
  {Miyasaka}}]{CoulonPRB07}%
  \BibitemOpen
  \bibfield  {author} {\bibinfo {author} {\bibfnamefont {C.}~\bibnamefont
  {Coulon}}, \bibinfo {author} {\bibfnamefont {R.}~\bibnamefont {Cl\'erac}},
  \bibinfo {author} {\bibfnamefont {W.}~\bibnamefont {Wernsdorfer}}, \bibinfo
  {author} {\bibfnamefont {T.}~\bibnamefont {Colin}}, \bibinfo {author}
  {\bibfnamefont {A.}~\bibnamefont {Saitoh}}, \bibinfo {author} {\bibfnamefont
  {N.}~\bibnamefont {Motokawa}}, \ and\ \bibinfo {author} {\bibfnamefont
  {H.}~\bibnamefont {Miyasaka}},\ }\Doi {10.1103/PhysRevB.76.214422} {\bibfield
   {journal} {\bibinfo  {journal} {Phys. Rev. B},\ }\textbf {\bibinfo {volume}
  {76}},\ \bibinfo {pages} {214422} (\bibinfo {year} {2007})}\BibitemShut
  {NoStop}%
\bibitem [{\citenamefont {Vindigni}\ and\ \citenamefont
  {Pini}(2009)}]{VindigniJPCM09}%
  \BibitemOpen
  \bibfield  {author} {\bibinfo {author} {\bibfnamefont {A.}~\bibnamefont
  {Vindigni}}\ and\ \bibinfo {author} {\bibfnamefont {M.~G.}\ \bibnamefont
  {Pini}},\ }\Doi {10.1088/0953-8984/21/23/236007} {\bibfield  {journal}
  {\bibinfo  {journal} {J. Phys.: Condens. Matter},\ }\textbf {\bibinfo
  {volume} {21}},\ \bibinfo {pages} {236007} (\bibinfo {year}
  {2009})}\BibitemShut {NoStop}%
\bibitem [{\citenamefont {Pini}\ \emph {et~al.}(2011)\citenamefont {Pini},
  \citenamefont {Rettori}, \citenamefont {Bogani}, \citenamefont {Lascialfari},
  \citenamefont {Mariani}, \citenamefont {Caneschi},\ and\ \citenamefont
  {Sessoli}}]{PiniPRB11}%
  \BibitemOpen
  \bibfield  {author} {\bibinfo {author} {\bibfnamefont {M.~G.}\ \bibnamefont
  {Pini}}, \bibinfo {author} {\bibfnamefont {A.}~\bibnamefont {Rettori}},
  \bibinfo {author} {\bibfnamefont {L.}~\bibnamefont {Bogani}}, \bibinfo
  {author} {\bibfnamefont {A.}~\bibnamefont {Lascialfari}}, \bibinfo {author}
  {\bibfnamefont {M.}~\bibnamefont {Mariani}}, \bibinfo {author} {\bibfnamefont
  {A.}~\bibnamefont {Caneschi}}, \ and\ \bibinfo {author} {\bibfnamefont
  {R.}~\bibnamefont {Sessoli}},\ }\Doi {10.1103/PhysRevB.84.094444} {\bibfield
  {journal} {\bibinfo  {journal} {Phys. Rev. B},\ }\textbf {\bibinfo {volume}
  {84}},\ \bibinfo {pages} {094444} (\bibinfo {year} {2011})}\BibitemShut
  {NoStop}%
\bibitem [{\citenamefont {Heintze}\ \emph {et~al.}(2013)\citenamefont
  {Heintze}, \citenamefont {Hallak}, \citenamefont {Clauss}, \citenamefont
  {Rettori}, \citenamefont {Pini}, \citenamefont {Totti}, \citenamefont
  {Dressel},\ and\ \citenamefont {Bogani}}]{Bogani_NatMat13}%
  \BibitemOpen
  \bibfield  {author} {\bibinfo {author} {\bibfnamefont {E.}~\bibnamefont
  {Heintze}}, \bibinfo {author} {\bibfnamefont {F.~E.}\ \bibnamefont {Hallak}},
  \bibinfo {author} {\bibfnamefont {C.}~\bibnamefont {Clauss}}, \bibinfo
  {author} {\bibfnamefont {A.}~\bibnamefont {Rettori}}, \bibinfo {author}
  {\bibfnamefont {M.~G.}\ \bibnamefont {Pini}}, \bibinfo {author}
  {\bibfnamefont {F.}~\bibnamefont {Totti}}, \bibinfo {author} {\bibfnamefont
  {M.}~\bibnamefont {Dressel}}, \ and\ \bibinfo {author} {\bibfnamefont
  {L.}~\bibnamefont {Bogani}},\ }\Doi {doi:10.1038/nmat3498} {\bibfield
  {journal} {\bibinfo  {journal} {Nature Mater.},\ }\textbf {\bibinfo {volume}
  {12}},\ \bibinfo {pages} {202} (\bibinfo {year} {2013})}\BibitemShut
  {NoStop}%
\bibitem [{\citenamefont {Caneschi}\ \emph {et~al.}(1991)\citenamefont
  {Caneschi}, \citenamefont {Gatteschi}, \citenamefont {Rey},\ and\
  \citenamefont {Sessoli}}]{caneschi1991}%
  \BibitemOpen
  \bibfield  {author} {\bibinfo {author} {\bibfnamefont {A.}~\bibnamefont
  {Caneschi}}, \bibinfo {author} {\bibfnamefont {D.}~\bibnamefont {Gatteschi}},
  \bibinfo {author} {\bibfnamefont {P.}~\bibnamefont {Rey}}, \ and\ \bibinfo
  {author} {\bibfnamefont {R.}~\bibnamefont {Sessoli}},\ }\href@noop {}
  {\bibfield  {journal} {\bibinfo  {journal} {Inorganic chemistry},\ }\textbf
  {\bibinfo {volume} {30}},\ \bibinfo {pages} {3936} (\bibinfo {year}
  {1991})}\BibitemShut {NoStop}%
\bibitem [{\citenamefont {Sessoli}\ \emph {et~al.}(2014)\citenamefont
  {Sessoli}, \citenamefont {Boulon}, \citenamefont {Caneschi}, \citenamefont
  {Mannini}, \citenamefont {Poggini}, \citenamefont {Wilhelm},\ and\
  \citenamefont {Rogalev}}]{Sessoli_NatPhys2014}%
  \BibitemOpen
  \bibfield  {author} {\bibinfo {author} {\bibfnamefont {R.}~\bibnamefont
  {Sessoli}}, \bibinfo {author} {\bibfnamefont {M.-E.}\ \bibnamefont {Boulon}},
  \bibinfo {author} {\bibfnamefont {A.}~\bibnamefont {Caneschi}}, \bibinfo
  {author} {\bibfnamefont {M.}~\bibnamefont {Mannini}}, \bibinfo {author}
  {\bibfnamefont {L.}~\bibnamefont {Poggini}}, \bibinfo {author} {\bibfnamefont
  {F.}~\bibnamefont {Wilhelm}}, \ and\ \bibinfo {author} {\bibfnamefont
  {A.}~\bibnamefont {Rogalev}},\ }\Doi {10.1038/nphys3152} {\bibfield
  {journal} {\bibinfo  {journal} {Nature Phys.},\ }\textbf {\bibinfo {volume}
  {10}} (\bibinfo {year} {2014})},\ \doi {10.1038/nphys3152}\BibitemShut
  {NoStop}%
\bibitem [{\citenamefont {Perdew}\ \emph {et~al.}(1996)\citenamefont {Perdew},
  \citenamefont {Burke},\ and\ \citenamefont {Ernzerhof}}]{perdew1996}%
  \BibitemOpen
  \bibfield  {author} {\bibinfo {author} {\bibfnamefont {J.~P.}\ \bibnamefont
  {Perdew}}, \bibinfo {author} {\bibfnamefont {K.}~\bibnamefont {Burke}}, \
  and\ \bibinfo {author} {\bibfnamefont {M.}~\bibnamefont {Ernzerhof}},\ }\Doi
  {10.1103/PhysRevLett.77.3865} {\bibfield  {journal} {\bibinfo  {journal}
  {Phys. Rev. Lett.},\ }\textbf {\bibinfo {volume} {77}},\ \bibinfo {pages}
  {3865} (\bibinfo {year} {1996})}\BibitemShut {NoStop}%
\bibitem [{\citenamefont {Kresse}\ and\ \citenamefont
  {Furthm\"uller}(1996)}]{kresse1996}%
  \BibitemOpen
  \bibfield  {author} {\bibinfo {author} {\bibfnamefont {G.}~\bibnamefont
  {Kresse}}\ and\ \bibinfo {author} {\bibfnamefont {J.}~\bibnamefont
  {Furthm\"uller}},\ }\Doi {10.1103/PhysRevB.54.11169} {\bibfield  {journal}
  {\bibinfo  {journal} {Phys. Rev. B},\ }\textbf {\bibinfo {volume} {54}},\
  \bibinfo {pages} {11169} (\bibinfo {year} {1996})}\BibitemShut {NoStop}%
\bibitem [{\citenamefont {Kresse}\ and\ \citenamefont
  {Joubert}(1999)}]{kresse1999}%
  \BibitemOpen
  \bibfield  {author} {\bibinfo {author} {\bibfnamefont {G.}~\bibnamefont
  {Kresse}}\ and\ \bibinfo {author} {\bibfnamefont {D.}~\bibnamefont
  {Joubert}},\ }\Doi {10.1103/PhysRevB.59.1758} {\bibfield  {journal} {\bibinfo
   {journal} {Phys. Rev. B},\ }\textbf {\bibinfo {volume} {59}},\ \bibinfo
  {pages} {1758} (\bibinfo {year} {1999})}\BibitemShut {NoStop}%
\bibitem [{\citenamefont {Bl\"ochl}(1994)}]{blochl1994}%
  \BibitemOpen
  \bibfield  {author} {\bibinfo {author} {\bibfnamefont {P.~E.}\ \bibnamefont
  {Bl\"ochl}},\ }\Doi {10.1103/PhysRevB.50.17953} {\bibfield  {journal}
  {\bibinfo  {journal} {Phys. Rev. B},\ }\textbf {\bibinfo {volume} {50}},\
  \bibinfo {pages} {17953} (\bibinfo {year} {1994})}\BibitemShut {NoStop}%
\bibitem [{\citenamefont {Hobbs}\ \emph {et~al.}(2000)\citenamefont {Hobbs},
  \citenamefont {Kresse},\ and\ \citenamefont {Hafner}}]{hobbs2000}%
  \BibitemOpen
  \bibfield  {author} {\bibinfo {author} {\bibfnamefont {D.}~\bibnamefont
  {Hobbs}}, \bibinfo {author} {\bibfnamefont {G.}~\bibnamefont {Kresse}}, \
  and\ \bibinfo {author} {\bibfnamefont {J.}~\bibnamefont {Hafner}},\ }\Doi
  {10.1103/PhysRevB.62.11556} {\bibfield  {journal} {\bibinfo  {journal} {Phys.
  Rev. B},\ }\textbf {\bibinfo {volume} {62}},\ \bibinfo {pages} {11556}
  (\bibinfo {year} {2000})}\BibitemShut {NoStop}%
\bibitem [{\citenamefont {Liechtenstein}\ \emph {et~al.}(1995)\citenamefont
  {Liechtenstein}, \citenamefont {Anisimov},\ and\ \citenamefont
  {Zaanen}}]{liechtenstein1995}%
  \BibitemOpen
  \bibfield  {author} {\bibinfo {author} {\bibfnamefont {A.~I.}\ \bibnamefont
  {Liechtenstein}}, \bibinfo {author} {\bibfnamefont {V.~I.}\ \bibnamefont
  {Anisimov}}, \ and\ \bibinfo {author} {\bibfnamefont {J.}~\bibnamefont
  {Zaanen}},\ }\Doi {10.1103/PhysRevB.52.R5467} {\bibfield  {journal} {\bibinfo
   {journal} {Phys. Rev. B},\ }\textbf {\bibinfo {volume} {52}},\ \bibinfo
  {pages} {R5467} (\bibinfo {year} {1995})}\BibitemShut {NoStop}%
\bibitem [{\citenamefont {Dudarev}\ \emph {et~al.}(1998)\citenamefont
  {Dudarev}, \citenamefont {Botton}, \citenamefont {Savrasov}, \citenamefont
  {Humphreys},\ and\ \citenamefont {Sutton}}]{dudarev1998}%
  \BibitemOpen
  \bibfield  {author} {\bibinfo {author} {\bibfnamefont {S.~L.}\ \bibnamefont
  {Dudarev}}, \bibinfo {author} {\bibfnamefont {G.~A.}\ \bibnamefont {Botton}},
  \bibinfo {author} {\bibfnamefont {S.~Y.}\ \bibnamefont {Savrasov}}, \bibinfo
  {author} {\bibfnamefont {C.~J.}\ \bibnamefont {Humphreys}}, \ and\ \bibinfo
  {author} {\bibfnamefont {A.~P.}\ \bibnamefont {Sutton}},\ }\Doi
  {10.1103/PhysRevB.57.1505} {\bibfield  {journal} {\bibinfo  {journal} {Phys.
  Rev. B},\ }\textbf {\bibinfo {volume} {57}},\ \bibinfo {pages} {1505}
  (\bibinfo {year} {1998})}\BibitemShut {NoStop}%
\bibitem [{\citenamefont {Xiang}\ \emph {et~al.}(2013)\citenamefont {Xiang},
  \citenamefont {Lee}, \citenamefont {Koo}, \citenamefont {Gong},\ and\
  \citenamefont {Whangbo}}]{whangbo2013}%
  \BibitemOpen
  \bibfield  {author} {\bibinfo {author} {\bibfnamefont {H.}~\bibnamefont
  {Xiang}}, \bibinfo {author} {\bibfnamefont {C.}~\bibnamefont {Lee}}, \bibinfo
  {author} {\bibfnamefont {H.-J.}\ \bibnamefont {Koo}}, \bibinfo {author}
  {\bibfnamefont {X.}~\bibnamefont {Gong}}, \ and\ \bibinfo {author}
  {\bibfnamefont {M.-H.}\ \bibnamefont {Whangbo}},\ }\href@noop {} {\bibfield
  {journal} {\bibinfo  {journal} {Dalton Trans.},\ }\textbf {\bibinfo {volume}
  {42}},\ \bibinfo {pages} {823} (\bibinfo {year} {2013})}\BibitemShut
  {NoStop}%
\bibitem [{\citenamefont {Wei}\ \emph {et~al.}(2005)\citenamefont {Wei},
  \citenamefont {Wang},\ and\ \citenamefont {Chen}}]{wei2005}%
  \BibitemOpen
  \bibfield  {author} {\bibinfo {author} {\bibfnamefont {H.}~\bibnamefont
  {Wei}}, \bibinfo {author} {\bibfnamefont {F.}~\bibnamefont {Wang}}, \ and\
  \bibinfo {author} {\bibfnamefont {Z.}~\bibnamefont {Chen}},\ }\href@noop {}
  {\bibfield  {journal} {\bibinfo  {journal} {Science in China Ser. B
  Chemistry},\ }\textbf {\bibinfo {volume} {48}},\ \bibinfo {pages} {402}
  (\bibinfo {year} {2005})}\BibitemShut {NoStop}%
\bibitem [{\citenamefont {Bartolom\'e}\ \emph {et~al.}(1996)\citenamefont
  {Bartolom\'e}, \citenamefont {Bartolom\'e}, \citenamefont {Benelli},
  \citenamefont {Caneschi}, \citenamefont {Gatteschi}, \citenamefont {Paulsen},
  \citenamefont {Pini}, \citenamefont {Rettori}, \citenamefont {Sessoli},\ and\
  \citenamefont {Volokitin}}]{Bartolome1996}%
  \BibitemOpen
  \bibfield  {author} {\bibinfo {author} {\bibfnamefont {F.}~\bibnamefont
  {Bartolom\'e}}, \bibinfo {author} {\bibfnamefont {J.}~\bibnamefont
  {Bartolom\'e}}, \bibinfo {author} {\bibfnamefont {C.}~\bibnamefont
  {Benelli}}, \bibinfo {author} {\bibfnamefont {A.}~\bibnamefont {Caneschi}},
  \bibinfo {author} {\bibfnamefont {D.}~\bibnamefont {Gatteschi}}, \bibinfo
  {author} {\bibfnamefont {C.}~\bibnamefont {Paulsen}}, \bibinfo {author}
  {\bibfnamefont {M.}~\bibnamefont {Pini}}, \bibinfo {author} {\bibfnamefont
  {A.}~\bibnamefont {Rettori}}, \bibinfo {author} {\bibfnamefont
  {R.}~\bibnamefont {Sessoli}}, \ and\ \bibinfo {author} {\bibfnamefont
  {Y.}~\bibnamefont {Volokitin}},\ }\Doi {10.1103/PhysRevLett.77.382}
  {\bibfield  {journal} {\bibinfo  {journal} {Phys. Rev. Lett.},\ }\textbf
  {\bibinfo {volume} {77}},\ \bibinfo {pages} {382} (\bibinfo {year}
  {1996})}\BibitemShut {NoStop}%
\bibitem [{\citenamefont {Caneschi}\ \emph
  {et~al.}(2002){\natexlab{a}}\citenamefont {Caneschi}, \citenamefont
  {Gatteschi}, \citenamefont {Lalioti}, \citenamefont {Sessoli}, \citenamefont
  {Sorace}, \citenamefont {Tangoulis},\ and\ \citenamefont
  {Vindigni}}]{caneschi2002}%
  \BibitemOpen
  \bibfield  {author} {\bibinfo {author} {\bibfnamefont {A.}~\bibnamefont
  {Caneschi}}, \bibinfo {author} {\bibfnamefont {D.}~\bibnamefont {Gatteschi}},
  \bibinfo {author} {\bibfnamefont {N.}~\bibnamefont {Lalioti}}, \bibinfo
  {author} {\bibfnamefont {R.}~\bibnamefont {Sessoli}}, \bibinfo {author}
  {\bibfnamefont {L.}~\bibnamefont {Sorace}}, \bibinfo {author} {\bibfnamefont
  {V.}~\bibnamefont {Tangoulis}}, \ and\ \bibinfo {author} {\bibfnamefont
  {A.}~\bibnamefont {Vindigni}},\ }\href@noop {} {\bibfield  {journal}
  {\bibinfo  {journal} {Chemistry-A European Journal},\ }\textbf {\bibinfo
  {volume} {8}},\ \bibinfo {pages} {286} (\bibinfo {year}
  {2002}{\natexlab{a}})}\BibitemShut {NoStop}%
\bibitem [{\citenamefont {Caneschi}\ \emph
  {et~al.}(2002){\natexlab{b}}\citenamefont {Caneschi}, \citenamefont
  {Gatteschi}, \citenamefont {Lalioti}, \citenamefont {Sangregorio},
  \citenamefont {Sessoli}, \citenamefont {Venturi}, \citenamefont {Vindigni},
  \citenamefont {Rettori}, \citenamefont {Pini},\ and\ \citenamefont
  {Novak}}]{CaneschiEPL2002}%
  \BibitemOpen
  \bibfield  {author} {\bibinfo {author} {\bibfnamefont {A.}~\bibnamefont
  {Caneschi}}, \bibinfo {author} {\bibfnamefont {D.}~\bibnamefont {Gatteschi}},
  \bibinfo {author} {\bibfnamefont {N.}~\bibnamefont {Lalioti}}, \bibinfo
  {author} {\bibfnamefont {C.}~\bibnamefont {Sangregorio}}, \bibinfo {author}
  {\bibfnamefont {R.}~\bibnamefont {Sessoli}}, \bibinfo {author} {\bibfnamefont
  {G.}~\bibnamefont {Venturi}}, \bibinfo {author} {\bibfnamefont
  {A.}~\bibnamefont {Vindigni}}, \bibinfo {author} {\bibfnamefont
  {A.}~\bibnamefont {Rettori}}, \bibinfo {author} {\bibfnamefont {M.~G.}\
  \bibnamefont {Pini}}, \ and\ \bibinfo {author} {\bibfnamefont {M.~A.}\
  \bibnamefont {Novak}},\ }\Doi {10.1209/epl/i2002-00416-x} {\bibfield
  {journal} {\bibinfo  {journal} {Europhys. Lett.},\ }\textbf {\bibinfo
  {volume} {58}},\ \bibinfo {pages} {771} (\bibinfo {year}
  {2002}{\natexlab{b}})}\BibitemShut {NoStop}%
\bibitem [{\citenamefont {Sangiorgio}\ \emph {et~al.}(2014)\citenamefont
  {Sangiorgio}, \citenamefont {Michaels}, \citenamefont {Pescia},\ and\
  \citenamefont {Vindigni}}]{Sangiorgio2014}%
  \BibitemOpen
  \bibfield  {author} {\bibinfo {author} {\bibfnamefont {B.}~\bibnamefont
  {Sangiorgio}}, \bibinfo {author} {\bibfnamefont {T.~C.~T.}\ \bibnamefont
  {Michaels}}, \bibinfo {author} {\bibfnamefont {D.}~\bibnamefont {Pescia}}, \
  and\ \bibinfo {author} {\bibfnamefont {A.}~\bibnamefont {Vindigni}},\ }\Doi
  {10.1103/PhysRevB.89.014429} {\bibfield  {journal} {\bibinfo  {journal}
  {Phys. Rev. B},\ }\textbf {\bibinfo {volume} {89}},\ \bibinfo {pages}
  {014429} (\bibinfo {year} {2014})}\BibitemShut {NoStop}%
\bibitem [{\citenamefont {Billoni}\ \emph {et~al.}(2011)\citenamefont
  {Billoni}, \citenamefont {Pianet}, \citenamefont {Pescia},\ and\
  \citenamefont {Vindigni}}]{Billoni2011}%
  \BibitemOpen
  \bibfield  {author} {\bibinfo {author} {\bibfnamefont {O.~V.}\ \bibnamefont
  {Billoni}}, \bibinfo {author} {\bibfnamefont {V.}~\bibnamefont {Pianet}},
  \bibinfo {author} {\bibfnamefont {D.}~\bibnamefont {Pescia}}, \ and\ \bibinfo
  {author} {\bibfnamefont {A.}~\bibnamefont {Vindigni}},\ }\Doi
  {10.1103/PhysRevB.84.064415} {\bibfield  {journal} {\bibinfo  {journal}
  {Phys. Rev. B},\ }\textbf {\bibinfo {volume} {84}},\ \bibinfo {pages}
  {064415} (\bibinfo {year} {2011})}\BibitemShut {NoStop}%
\bibitem [{\citenamefont {{Yan}}\ and\ \citenamefont {{Bauer}}(2011)}]{Yan_12}%
  \BibitemOpen
  \bibfield  {author} {\bibinfo {author} {\bibfnamefont {P.}~\bibnamefont
  {{Yan}}}\ and\ \bibinfo {author} {\bibfnamefont {G.~E.~W.}\ \bibnamefont
  {{Bauer}}},\ }\Doi {10.1103/PhysRevLett.109.087202} {\bibfield  {journal}
  {\bibinfo  {journal} {Phys. Rev. Lett.},\ }\textbf {\bibinfo {volume}
  {109}},\ \bibinfo {pages} {087202} (\bibinfo {year} {2011})}\BibitemShut
  {NoStop}%
\bibitem [{\citenamefont {{Fogedby}}\ \emph {et~al.}(1984)\citenamefont
  {{Fogedby}}, \citenamefont {{Hedegard}},\ and\ \citenamefont
  {{Svane}}}]{Fogedby84JPCSSP}%
  \BibitemOpen
  \bibfield  {author} {\bibinfo {author} {\bibfnamefont {H.~C.}\ \bibnamefont
  {{Fogedby}}}, \bibinfo {author} {\bibfnamefont {P.}~\bibnamefont
  {{Hedegard}}}, \ and\ \bibinfo {author} {\bibfnamefont {A.}~\bibnamefont
  {{Svane}}},\ }\href@noop {} {\bibfield  {journal} {\bibinfo  {journal} {J.
  Phys. C: Solid State Phys.},\ }\textbf {\bibinfo {volume} {17}},\ \bibinfo
  {pages} {3475} (\bibinfo {year} {1984})}\BibitemShut {NoStop}%
\bibitem [{\citenamefont {Braun}(1994)}]{HB_Braun_PRB94}%
  \BibitemOpen
  \bibfield  {author} {\bibinfo {author} {\bibfnamefont {H.-B.}\ \bibnamefont
  {Braun}},\ }\Doi {10.1103/PhysRevB.50.16485} {\bibfield  {journal} {\bibinfo
  {journal} {Phys. Rev. B},\ }\textbf {\bibinfo {volume} {50}},\ \bibinfo
  {pages} {16485} (\bibinfo {year} {1994})}\BibitemShut {NoStop}%
\bibitem [{\citenamefont {Enz}(1964)}]{Enz}%
  \BibitemOpen
  \bibfield  {author} {\bibinfo {author} {\bibfnamefont {U.}~\bibnamefont
  {Enz}},\ }\Doi {http://dx.doi.org/10.5169/seals-113481} {\bibfield  {journal}
  {\bibinfo  {journal} {Helv. phys. Acta},\ }\textbf {\bibinfo {volume} {37}},\
  \bibinfo {pages} {245} (\bibinfo {year} {1964})}\BibitemShut {NoStop}%
\bibitem [{\citenamefont {Seiden}(1983)}]{Seiden}%
  \BibitemOpen
  \bibfield  {author} {\bibinfo {author} {\bibfnamefont {J.}~\bibnamefont
  {Seiden}},\ }\Doi {10.1051/jphyslet:019830044023094700} {\bibfield  {journal}
  {\bibinfo  {journal} {J. Phys. Lett. (Paris)},\ }\textbf {\bibinfo {volume}
  {44}},\ \bibinfo {pages} {947} (\bibinfo {year} {1983})}\BibitemShut
  {NoStop}%
\bibitem [{\citenamefont {Vindigni}\ \emph {et~al.}(2004)\citenamefont
  {Vindigni}, \citenamefont {Regnault},\ and\ \citenamefont
  {Jolicoeur}}]{VindigniPRB04}%
  \BibitemOpen
  \bibfield  {author} {\bibinfo {author} {\bibfnamefont {A.}~\bibnamefont
  {Vindigni}}, \bibinfo {author} {\bibfnamefont {N.}~\bibnamefont {Regnault}},
  \ and\ \bibinfo {author} {\bibfnamefont {T.}~\bibnamefont {Jolicoeur}},\
  }\Doi {10.1103/PhysRevB.70.134423} {\bibfield  {journal} {\bibinfo  {journal}
  {Phys. Rev. B},\ }\textbf {\bibinfo {volume} {70}},\ \bibinfo {pages}
  {134423} (\bibinfo {year} {2004})}\BibitemShut {NoStop}%
\bibitem [{\citenamefont {{Tretiakov}}\ and\ \citenamefont
  {{Abanov}}(2010)}]{Ar_Abanov_PRL_10a}%
  \BibitemOpen
  \bibfield  {author} {\bibinfo {author} {\bibfnamefont {O.~A.}\ \bibnamefont
  {{Tretiakov}}}\ and\ \bibinfo {author} {\bibfnamefont {A.}~\bibnamefont
  {{Abanov}}},\ }\Doi {10.1103/PhysRevLett.105.157201} {\bibfield  {journal}
  {\bibinfo  {journal} {Phys. Rev. Lett.},\ }\textbf {\bibinfo {volume}
  {105}},\ \bibinfo {pages} {157201} (\bibinfo {year} {2010})}\BibitemShut
  {NoStop}%
\bibitem [{\citenamefont {{Tretiakov}}\ \emph {et~al.}(2010)\citenamefont
  {{Tretiakov}}, \citenamefont {{Liu}},\ and\ \citenamefont
  {{Abanov}}}]{Ar_Abanov_PRL_10b}%
  \BibitemOpen
  \bibfield  {author} {\bibinfo {author} {\bibfnamefont {O.~A.}\ \bibnamefont
  {{Tretiakov}}}, \bibinfo {author} {\bibfnamefont {Y.}~\bibnamefont {{Liu}}},
  \ and\ \bibinfo {author} {\bibfnamefont {A.}~\bibnamefont {{Abanov}}},\ }\Doi
  {10.1103/PhysRevLett.105.217203} {\bibfield  {journal} {\bibinfo  {journal}
  {Phys. Rev. Lett.},\ }\textbf {\bibinfo {volume} {105}},\ \bibinfo {pages}
  {217203} (\bibinfo {year} {2010})}\BibitemShut {NoStop}%
\bibitem [{\citenamefont {Braun}(2012)}]{B_Braun_AdvPhys_2012}%
  \BibitemOpen
  \bibfield  {author} {\bibinfo {author} {\bibfnamefont {H.-B.}\ \bibnamefont
  {Braun}},\ }\Doi {10.1080/00018732.2012.663070} {\bibfield  {journal}
  {\bibinfo  {journal} {Adv. Phys.},\ }\textbf {\bibinfo {volume} {61}},\
  \bibinfo {pages} {1} (\bibinfo {year} {2012})}\BibitemShut {NoStop}%
\bibitem [{\citenamefont {{Parkin}}\ \emph {et~al.}(2008)\citenamefont
  {{Parkin}}, \citenamefont {{Hayashi}},\ and\ \citenamefont
  {{Thomas}}}]{Parkin_Science_08}%
  \BibitemOpen
  \bibfield  {author} {\bibinfo {author} {\bibfnamefont {S.~S.~P.}\
  \bibnamefont {{Parkin}}}, \bibinfo {author} {\bibfnamefont {M.}~\bibnamefont
  {{Hayashi}}}, \ and\ \bibinfo {author} {\bibfnamefont {L.}~\bibnamefont
  {{Thomas}}},\ }\Doi {10.1126/science.1145799} {\bibfield  {journal} {\bibinfo
   {journal} {Science},\ }\textbf {\bibinfo {volume} {320}},\ \bibinfo {pages}
  {190} (\bibinfo {year} {2008})}\BibitemShut {NoStop}%
\bibitem [{\citenamefont {{Hayashi}}\ \emph {et~al.}(2008)\citenamefont
  {{Hayashi}}, \citenamefont {{Thomas}}, \citenamefont {{Moriya}},
  \citenamefont {{Rettner}},\ and\ \citenamefont
  {{Parkin}}}]{Hayashi_Science_08}%
  \BibitemOpen
  \bibfield  {author} {\bibinfo {author} {\bibfnamefont {M.}~\bibnamefont
  {{Hayashi}}}, \bibinfo {author} {\bibfnamefont {L.}~\bibnamefont {{Thomas}}},
  \bibinfo {author} {\bibfnamefont {R.}~\bibnamefont {{Moriya}}}, \bibinfo
  {author} {\bibfnamefont {C.}~\bibnamefont {{Rettner}}}, \ and\ \bibinfo
  {author} {\bibfnamefont {S.~S.~P.}\ \bibnamefont {{Parkin}}},\ }\Doi
  {10.1126/science.1154587} {\bibfield  {journal} {\bibinfo  {journal}
  {Science},\ }\textbf {\bibinfo {volume} {320}},\ \bibinfo {pages} {209}
  (\bibinfo {year} {2008})}\BibitemShut {NoStop}%
\bibitem [{\citenamefont {{Vanhaverbeke}}\ \emph {et~al.}(2008)\citenamefont
  {{Vanhaverbeke}}, \citenamefont {{Bischof}},\ and\ \citenamefont
  {{Allenspach}}}]{Allenspach_PRL08}%
  \BibitemOpen
  \bibfield  {author} {\bibinfo {author} {\bibfnamefont {A.}~\bibnamefont
  {{Vanhaverbeke}}}, \bibinfo {author} {\bibfnamefont {A.}~\bibnamefont
  {{Bischof}}}, \ and\ \bibinfo {author} {\bibfnamefont {R.}~\bibnamefont
  {{Allenspach}}},\ }\Doi {10.1103/PhysRevLett.101.107202} {\bibfield
  {journal} {\bibinfo  {journal} {Phys. Rev. Lett.},\ }\textbf {\bibinfo
  {volume} {101}},\ \bibinfo {pages} {107202} (\bibinfo {year}
  {2008})}\BibitemShut {NoStop}%
\bibitem [{\citenamefont {{Tatara}}\ and\ \citenamefont
  {{Kohno}}(2004)}]{Tatara_PRL04}%
  \BibitemOpen
  \bibfield  {author} {\bibinfo {author} {\bibfnamefont {G.}~\bibnamefont
  {{Tatara}}}\ and\ \bibinfo {author} {\bibfnamefont {H.}~\bibnamefont
  {{Kohno}}},\ }\Doi {10.1103/PhysRevLett.92.086601} {\bibfield  {journal}
  {\bibinfo  {journal} {Phys. Rev. Lett.},\ }\textbf {\bibinfo {volume} {92}},\
  \bibinfo {pages} {086601} (\bibinfo {year} {2004})}\BibitemShut {NoStop}%
\bibitem [{\citenamefont {{Yan}}\ \emph {et~al.}(2011)\citenamefont {{Yan}},
  \citenamefont {{Wang}},\ and\ \citenamefont {{Wang}}}]{Yan_11}%
  \BibitemOpen
  \bibfield  {author} {\bibinfo {author} {\bibfnamefont {P.}~\bibnamefont
  {{Yan}}}, \bibinfo {author} {\bibfnamefont {X.~S.}\ \bibnamefont {{Wang}}}, \
  and\ \bibinfo {author} {\bibfnamefont {X.~R.}\ \bibnamefont {{Wang}}},\ }\Doi
  {10.1103/PhysRevLett.107.177207} {\bibfield  {journal} {\bibinfo  {journal}
  {Phys. Rev. Lett.},\ }\textbf {\bibinfo {volume} {107}},\ \bibinfo {pages}
  {177207} (\bibinfo {year} {2011})}\BibitemShut {NoStop}%
\bibitem [{\citenamefont {Wagni\`ere}(1989)}]{Wagniere_PRA_89}%
  \BibitemOpen
  \bibfield  {author} {\bibinfo {author} {\bibfnamefont {G.}~\bibnamefont
  {Wagni\`ere}},\ }\Doi {10.1103/PhysRevA.40.2437} {\bibfield  {journal}
  {\bibinfo  {journal} {Phys. Rev. A},\ }\textbf {\bibinfo {volume} {40}},\
  \bibinfo {pages} {2437} (\bibinfo {year} {1989})}\BibitemShut {NoStop}%
\bibitem [{\citenamefont {G\"ohler}\ \emph {et~al.}(2011)\citenamefont
  {G\"ohler}, \citenamefont {Hamelbeck}, \citenamefont {Markus}, \citenamefont
  {Kettner}, \citenamefont {Hanne}, \citenamefont {Vager}, \citenamefont
  {Naaman},\ and\ \citenamefont {Zacharias}}]{Goehler_Science11}%
  \BibitemOpen
  \bibfield  {author} {\bibinfo {author} {\bibfnamefont {B.}~\bibnamefont
  {G\"ohler}}, \bibinfo {author} {\bibfnamefont {V.}~\bibnamefont {Hamelbeck}},
  \bibinfo {author} {\bibfnamefont {T.~Z.}\ \bibnamefont {Markus}}, \bibinfo
  {author} {\bibfnamefont {M.}~\bibnamefont {Kettner}}, \bibinfo {author}
  {\bibfnamefont {G.~F.}\ \bibnamefont {Hanne}}, \bibinfo {author}
  {\bibfnamefont {Z.}~\bibnamefont {Vager}}, \bibinfo {author} {\bibfnamefont
  {R.}~\bibnamefont {Naaman}}, \ and\ \bibinfo {author} {\bibfnamefont
  {H.}~\bibnamefont {Zacharias}},\ }\Doi {10.1126/science.1199339} {\bibfield
  {journal} {\bibinfo  {journal} {Science},\ }\textbf {\bibinfo {volume}
  {331}},\ \bibinfo {pages} {894} (\bibinfo {year} {2011})}\BibitemShut
  {NoStop}%
\bibitem [{\citenamefont {Manso}\ \emph {et~al.}(2014)\citenamefont {Manso},
  \citenamefont {Chai}, \citenamefont {Atack}, \citenamefont {Furi},
  \citenamefont {De~Ste~Croix}, \citenamefont {Haigh}, \citenamefont
  {Trappetti}, \citenamefont {Ogunniyi}, \citenamefont {Shewell}, \citenamefont
  {Boitano}, \citenamefont {Clark}, \citenamefont {Korlach}, \citenamefont
  {Blades}, \citenamefont {Mirkes}, \citenamefont {Gorban}, \citenamefont
  {Paton}, \citenamefont {Jennings},\ and\ \citenamefont
  {Oggioni}}]{Manso_NatComm14}%
  \BibitemOpen
  \bibfield  {author} {\bibinfo {author} {\bibfnamefont {A.~S.}\ \bibnamefont
  {Manso}}, \bibinfo {author} {\bibfnamefont {M.~H.}\ \bibnamefont {Chai}},
  \bibinfo {author} {\bibfnamefont {J.~M.}\ \bibnamefont {Atack}}, \bibinfo
  {author} {\bibfnamefont {L.}~\bibnamefont {Furi}}, \bibinfo {author}
  {\bibfnamefont {M.}~\bibnamefont {De~Ste~Croix}}, \bibinfo {author}
  {\bibfnamefont {R.}~\bibnamefont {Haigh}}, \bibinfo {author} {\bibfnamefont
  {C.}~\bibnamefont {Trappetti}}, \bibinfo {author} {\bibfnamefont {A.~D.}\
  \bibnamefont {Ogunniyi}}, \bibinfo {author} {\bibfnamefont {L.~K.}\
  \bibnamefont {Shewell}}, \bibinfo {author} {\bibfnamefont {M.}~\bibnamefont
  {Boitano}}, \bibinfo {author} {\bibfnamefont {T.~A.}\ \bibnamefont {Clark}},
  \bibinfo {author} {\bibfnamefont {J.}~\bibnamefont {Korlach}}, \bibinfo
  {author} {\bibfnamefont {M.}~\bibnamefont {Blades}}, \bibinfo {author}
  {\bibfnamefont {E.}~\bibnamefont {Mirkes}}, \bibinfo {author} {\bibfnamefont
  {A.~N.}\ \bibnamefont {Gorban}}, \bibinfo {author} {\bibfnamefont {J.~C.}\
  \bibnamefont {Paton}}, \bibinfo {author} {\bibfnamefont {M.~P.}\ \bibnamefont
  {Jennings}}, \ and\ \bibinfo {author} {\bibfnamefont {M.~R.}\ \bibnamefont
  {Oggioni}},\ }\Doi {10.1038/ncomms6055} {\bibfield  {journal} {\bibinfo
  {journal} {Nat. Commun.},\ }\textbf {\bibinfo {volume} {5}},\ \bibinfo
  {pages} {5055} (\bibinfo {year} {2014})}\BibitemShut {NoStop}%
\bibitem [{\citenamefont {Bode}\ \emph {et~al.}(2007)\citenamefont {Bode},
  \citenamefont {Heide}, \citenamefont {von Bergmann}, \citenamefont
  {Ferriani}, \citenamefont {Heinze}, \citenamefont {Bihlmayer}, \citenamefont
  {Kubetzka}, \citenamefont {Pietzsch}, \citenamefont {Bl\"ugel},\ and\
  \citenamefont {Wiesendanger}}]{Bode2007}%
  \BibitemOpen
  \bibfield  {author} {\bibinfo {author} {\bibfnamefont {M.}~\bibnamefont
  {Bode}}, \bibinfo {author} {\bibfnamefont {M.}~\bibnamefont {Heide}},
  \bibinfo {author} {\bibfnamefont {K.}~\bibnamefont {von Bergmann}}, \bibinfo
  {author} {\bibfnamefont {P.}~\bibnamefont {Ferriani}}, \bibinfo {author}
  {\bibfnamefont {S.}~\bibnamefont {Heinze}}, \bibinfo {author} {\bibfnamefont
  {G.}~\bibnamefont {Bihlmayer}}, \bibinfo {author} {\bibfnamefont
  {A.}~\bibnamefont {Kubetzka}}, \bibinfo {author} {\bibfnamefont
  {O.}~\bibnamefont {Pietzsch}}, \bibinfo {author} {\bibfnamefont
  {S.}~\bibnamefont {Bl\"ugel}}, \ and\ \bibinfo {author} {\bibfnamefont
  {R.}~\bibnamefont {Wiesendanger}},\ }\Doi {10.1038/nature05802} {\bibfield
  {journal} {\bibinfo  {journal} {Nature},\ }\textbf {\bibinfo {volume}
  {447}},\ \bibinfo {pages} {190} (\bibinfo {year} {2007})}\BibitemShut
  {NoStop}%
\bibitem [{\citenamefont {Blume}\ \emph {et~al.}(1975)\citenamefont {Blume},
  \citenamefont {Heller},\ and\ \citenamefont {Lurie}}]{Blume75PRB}%
  \BibitemOpen
  \bibfield  {author} {\bibinfo {author} {\bibfnamefont {M.}~\bibnamefont
  {Blume}}, \bibinfo {author} {\bibfnamefont {P.}~\bibnamefont {Heller}}, \
  and\ \bibinfo {author} {\bibfnamefont {N.~A.}\ \bibnamefont {Lurie}},\ }\Doi
  {10.1103/PhysRevB.11.4483} {\bibfield  {journal} {\bibinfo  {journal} {Phys.
  Rev. B},\ }\textbf {\bibinfo {volume} {11}},\ \bibinfo {pages} {4483}
  (\bibinfo {year} {1975})}\BibitemShut {NoStop}%
\bibitem [{\citenamefont {Pandit}\ and\ \citenamefont
  {Tannous}(1983)}]{Tannous}%
  \BibitemOpen
  \bibfield  {author} {\bibinfo {author} {\bibfnamefont {R.}~\bibnamefont
  {Pandit}}\ and\ \bibinfo {author} {\bibfnamefont {C.}~\bibnamefont
  {Tannous}},\ }\Doi {10.1103/PhysRevB.28.281} {\bibfield  {journal} {\bibinfo
  {journal} {Phys. Rev. B},\ }\textbf {\bibinfo {volume} {28}},\ \bibinfo
  {pages} {281} (\bibinfo {year} {1983})}\BibitemShut {NoStop}%
\bibitem [{\citenamefont {Wyld}(1976)}]{Wyld}%
  \BibitemOpen
  \bibfield  {author} {\bibinfo {author} {\bibfnamefont {H.~W.}\ \bibnamefont
  {Wyld}},\ }\href@noop {} {\emph {\bibinfo {title} {Mathematical Methods of
  Physics}}}\ (\bibinfo  {publisher} {Benjamin},\ \bibinfo {address}
  {Massachusetts, USA},\ \bibinfo {year} {1976})\BibitemShut {NoStop}%
\bibitem [{\citenamefont {Stroud}(1971)}]{Stroud}%
  \BibitemOpen
  \bibfield  {author} {\bibinfo {author} {\bibfnamefont {A.~H.}\ \bibnamefont
  {Stroud}},\ }\href@noop {} {\emph {\bibinfo {title} {Approximate Calculation
  of Multiple Integrals}}}\ (\bibinfo  {publisher} {Prentice-Hall, Englewood
  Cliffs},\ \bibinfo {address} {New Jersey, USA},\ \bibinfo {year}
  {1971})\BibitemShut {NoStop}%
\bibitem [{\citenamefont {McLaren}(1963)}]{McLaren}%
  \BibitemOpen
  \bibfield  {author} {\bibinfo {author} {\bibfnamefont {A.~D.}\ \bibnamefont
  {McLaren}},\ }\href@noop {} {\bibfield  {journal} {\bibinfo  {journal} {Math.
  Comp.},\ }\textbf {\bibinfo {volume} {17}},\ \bibinfo {pages} {361} (\bibinfo
  {year} {1963})}\BibitemShut {NoStop}%
\bibitem [{\citenamefont {Abramowitz}\ and\ \citenamefont
  {Stegum}(1970)}]{Abramowitz}%
  \BibitemOpen
  \bibfield  {author} {\bibinfo {author} {\bibfnamefont {M.}~\bibnamefont
  {Abramowitz}}\ and\ \bibinfo {author} {\bibfnamefont {I.~E.}\ \bibnamefont
  {Stegum}},\ }\href@noop {} {\emph {\bibinfo {title} {Handbook of mathematical
  functions}}}\ (\bibinfo  {publisher} {Dover},\ \bibinfo {address} {New York,
  USA},\ \bibinfo {year} {1970})\BibitemShut {NoStop}%
\bibitem [{\citenamefont {Vindigni}\ \emph {et~al.}(2006)\citenamefont
  {Vindigni}, \citenamefont {Rettori}, \citenamefont {Pini}, \citenamefont
  {Carbone},\ and\ \citenamefont {Gambardella}}]{Vindigni06APA}%
  \BibitemOpen
  \bibfield  {author} {\bibinfo {author} {\bibfnamefont {A.}~\bibnamefont
  {Vindigni}}, \bibinfo {author} {\bibfnamefont {A.}~\bibnamefont {Rettori}},
  \bibinfo {author} {\bibfnamefont {M.}~\bibnamefont {Pini}}, \bibinfo {author}
  {\bibfnamefont {C.}~\bibnamefont {Carbone}}, \ and\ \bibinfo {author}
  {\bibfnamefont {P.}~\bibnamefont {Gambardella}},\ }\Doi
  {10.1007/s00339-005-3364-4} {\bibfield  {journal} {\bibinfo  {journal} {Appl.
  Phys. A},\ }\textbf {\bibinfo {volume} {82}},\ \bibinfo {pages} {385}
  (\bibinfo {year} {2006})}\BibitemShut {NoStop}%
\bibitem [{\citenamefont {Hinzke}\ and\ \citenamefont
  {Nowak}(2000)}]{Hinzke00PRB}%
  \BibitemOpen
  \bibfield  {author} {\bibinfo {author} {\bibfnamefont {D.}~\bibnamefont
  {Hinzke}}\ and\ \bibinfo {author} {\bibfnamefont {U.}~\bibnamefont {Nowak}},\
  }\Doi {10.1103/PhysRevB.61.6734} {\bibfield  {journal} {\bibinfo  {journal}
  {Phys. Rev. B},\ }\textbf {\bibinfo {volume} {61}},\ \bibinfo {pages} {6734}
  (\bibinfo {year} {2000})}\BibitemShut {NoStop}%
\bibitem [{\citenamefont {da~Silva}\ \emph {et~al.}(1995)\citenamefont
  {da~Silva}, \citenamefont {Moreira}, \citenamefont {Soares},\ and\
  \citenamefont {Barreto}}]{daSilva_PRE_95}%
  \BibitemOpen
  \bibfield  {author} {\bibinfo {author} {\bibfnamefont {J.~K.~L.}\
  \bibnamefont {da~Silva}}, \bibinfo {author} {\bibfnamefont {A.~G.}\
  \bibnamefont {Moreira}}, \bibinfo {author} {\bibfnamefont {M.~S.}\
  \bibnamefont {Soares}}, \ and\ \bibinfo {author} {\bibfnamefont {F.~C.~S.}\
  \bibnamefont {Barreto}},\ }\Doi {10.1103/PhysRevE.52.4527} {\bibfield
  {journal} {\bibinfo  {journal} {Phys. Rev. E},\ }\textbf {\bibinfo {volume}
  {52}},\ \bibinfo {pages} {4527} (\bibinfo {year} {1995})}\BibitemShut
  {NoStop}%
\bibitem [{\citenamefont {Gilbert}(2004)}]{Gilbert_2004}%
  \BibitemOpen
  \bibfield  {author} {\bibinfo {author} {\bibfnamefont {T.}~\bibnamefont
  {Gilbert}},\ }\Doi {10.1109/TMAG.2004.836740} {\bibfield  {journal} {\bibinfo
   {journal} {Magnetics, IEEE Transactions on},\ }\textbf {\bibinfo {volume}
  {40}},\ \bibinfo {pages} {3443} (\bibinfo {year} {2004})},\ ISSN \bibinfo
  {issn} {0018-9464}\BibitemShut {NoStop}%
\end{thebibliography}

%

\end{document}